\documentclass[onecollarge]{svjour2}

\smartqed  

\usepackage[numbers,sort&compress]{natbib}

\def\makeheadbox{{}}

\usepackage{times}
\usepackage{txfonts}
\usepackage{graphicx}
\usepackage{comment}
\usepackage{color}
\usepackage{fancyhdr}
\usepackage{subfigure}
\usepackage{setspace}
\usepackage{array}
\usepackage{xspace}
\usepackage{upgreek}
\usepackage{multirow}

\usepackage{lineno}

\newcommand{\beq}{\begin{eqnarray}}
\newcommand{\eeq}{\end{eqnarray}}
\newcommand{\be}{\begin{eqnarray*}}
\newcommand{\ee}{\end{eqnarray*}}

\newcommand{\ct}[1]{{Table.~\ref{#1}}}

\def\lsim{\raise0.3ex\hbox{$<$\kern-0.75em\raise-1.1ex\hbox{$\sim$}}}
\def\gsim{\raise0.3ex\hbox{$>$\kern-0.75em\raise-1.1ex\hbox{$\sim$}}}

\def\AFTER {\mbox{AFTER@LHC}\xspace}
\def\pKr {$p$Kr\xspace}	
\def\pXe {$p$Xe\xspace}
\def\pp   {$pp$\xspace}

\def\pA   {$pA$\xspace}

\def\AA   {$AA$\xspace}
\def\AB   {$AB$\xspace}
\def\PbPb {PbPb\xspace}

\def\PbA {Pb$A$\xspace}
\def\PbXe {PbXe\xspace}
\def\pPb {$p$Pb\xspace}

\def\pXe {$p$Xe\xspace}
\def\pAr {$p$Ar\xspace}
\def\pNe {$p$Ne\xspace}
\def\Pbp {Pb$p$\xspace}
\def\pA {$pA$\xspace}

\def\beq     {\begin{equation}}
\def\eeq     {\end{equation}}

\def\jpsi   {\mbox{$J/\psi$}\xspace}
\newcommand{\psip}{\mbox{$\psi'$}\xspace}
\newcommand{\chic}{\mbox{$\chi_c$}\xspace}
\newcommand{\chib}{\mbox{$\chi_b$}\xspace}
\newcommand{\ups}{\mbox{$\Upsilon$}\xspace}
\newcommand{\bb}{\mbox{$b\bar{b}$}\xspace}
\newcommand{\cc}{\mbox{$c\bar{c}$}\xspace}

\def\pT      {\mbox{$p_{T}$}}

\def\beq     {\begin{equation}}
\def\eeq     {\end{equation}}

\def\sqrtsNN {\mbox{$\sqrt{s_{NN}}$}}

\def\RAA     {\mbox{$R_{AA}$}}

\def\ycms   {\mbox{$y_{\rm c.m.s.}$}\xspace}

\newcommand{\Pb}{{\rm Pb}}

\usepackage{ifpdf}
\ifpdf
\usepackage[pdftex]{hyperref}
\else
\usepackage[hypertex]{hyperref}
\fi

\hypersetup{
  pdftitle={Heavy-ion Physics at a Fixed-Target Experiment Using the LHC Proton and
Lead Beams (\AFTER): Feasibility Studies for Quarkonium and Drell-Yan Production},%
  pdfauthor={B. Trzeciak et al.},%
  pdfsubject={},%
  pdfkeywords={Quarkonium Production, Quark-Gluon Plasma, Fixed-Target experiment, Large Hadron Collider},%
  pdfstartview={},%
  bookmarksopen=true, breaklinks=true, debug=true, %
  colorlinks=true, linkcolor=red, citecolor=blue, urlcolor=blue
}

\journalname{Few Body Systems}
\begin{document}

\title{Heavy-ion Physics at a Fixed-Target Experiment Using the LHC Proton and
Lead Beams (\AFTER): Feasibility Studies for Quarkonium and Drell-Yan Production}

\author{B.~Trzeciak \and C. Da Silva \and E.G. Ferreiro \and C.~Hadjidakis \and D.~Kikola  \and J.P.~Lansberg \and L.~Massacrier \and J.~ Seixas \and A.~Uras \and Z.~Yang}

\institute{  
B.~Trzeciak \at Institute for Subatomic Physics, Utrecht University, Utrecht, The Netherlands\\
C.~Da Silva \at P-25, Los Alamos National Laboratory, Los Alamos, NM 87545, USA\\
E.G. Ferreiro \at Departamento de F{\'\i}sica de Part{\'\i}culas, Universidade de Santiago de Compostela, 15782 Santiago de Compostela, Spain \\
C.~Hadjidakis, J.P. Lansberg, L. Massacrier \at IPNO, CNRS-IN2P3, Univ. Paris-Sud, Universit\'e Paris-Saclay,  Orsay, France\\
D.~Kikola \at Faculty of Physics, Warsaw University of Technology,  Warsaw, Poland \\
J.~Seixas \at LIP and IST, Lisbon, Portugal\\
A.~Uras \at IPNL, Lyon, France\\
Z.~Yang \at Tsinghua University, Beijing, China
}

\date{}

\maketitle

\begin{abstract}
We outline the case for heavy-ion-physics studies using the multi-TeV lead LHC beams in the fixed-target mode. After a brief contextual reminder, we detail the possible contributions of \AFTER to heavy-ion physics with a specific emphasis on quarkonia. We then present  performance simulations for a selection of observables. These show that $\Upsilon(nS)$, $J/\psi$ and $\psi(2S)$ production in heavy-ion collisions can be studied in new energy and rapidity domains with the LHCb and ALICE detectors.
We also discuss the relevance to analyse the Drell-Yan pair production in asymmetric nucleus-nucleus collisions to study the factorisation of the nuclear modification of partonic densities and of further quarkonia to restore their status of golden probes of the quark-gluon plasma formation.
\end{abstract}

\tableofcontents

\section{Introduction}
\label{intro}
Hadro-production experiments in the fixed-target mode have indisputably played a fundamental role in the history of quarkonium physics. This began with the co-discovery of the $J/\psi$~\cite{Aubert:1974js} in 1974 at the Brookhaven National Laboratory and followed by the discovery of
the $\Upsilon$~\cite{Herb:1977ek} and the first observation of $h_c$~\cite{Armstrong:1992ae} at the Fermi National Laboratory. The dedicated heavy-ion program at the Super Proton Synchrotron (SPS) at CERN then uncovered many novel and unexpected features of quark and gluon dynamics, including the anomalous suppression of $J/\psi$~\cite{Abreu:2000ni}
in PbPb collisions, and helped discover the strong non-factorising nuclear suppression 
of the $J/\psi$ hadro-production at high $x_F$~\cite{Hoyer:1990us}. Let us recall that the first observation of $J/\psi$-pair production was carried out at the SPS as early as in 1982~\cite{Badier:1982ae}. Fixed-target experiments have also played a central role in measuring Drell-Yan pair production, in particular below the bottomonium region~\cite{Baldit:1994jk,Hawker:1998ty,Vasilev:1999fa,Zhu:2006gx}.

In this context, we find it important to remind that collisions of the proton and heavy-ion LHC
beams on fixed targets open a remarkably large range of physics opportunities~\cite{Brodsky:2012vg,Koshkarev:2016ket,Signori:2016lvd,Koshkarev:2016acq,Lansberg:2016gwm,Lansberg:2016urh,Signori:2016jwo,Pisano:2015fzm,Vogt:2015dva,Fengand:2015rka,Barschel:2015mka,Kikola:2015lka,Kurepinand:2015jka,Zhou:2015wea,Arleo:2015lja,Lansberg:2015lva,Brodsky:2015fna,Massacrier:2015qba,Anselmino:2015eoa,Lansberg:2015kha,Ceccopieri:2015rha,Goncalves:2015hra,Kanazawa:2015fia,Lansberg:2015hla,Massacrier:2015nsm,Lansberg:2014myg,Chen:2014hqa,Rakotozafindrabe:2013au,Lansberg:2013wpx,Lansberg:2012sq,Lorce:2012rn,Rakotozafindrabe:2012ei,Boer:2012bt,Lansberg:2012wj,Lansberg:2012kf,Liu:2012vn}. 
The \AFTER project bears on them and aims at an ambitious heavy-ion, spin and hadron-structure physics programme.
These opportunities result from a high luminosity -- be it with an internal gas target or with an extracted beam-- together with a relatively high center-of-mass system (c.m.s.) energy of 115 GeV per nucleon with a 7 TeV proton beam and 72 GeV per nucleon with a 2.76 A.TeV lead beam. 
Besides, the boost typical of the fixed-target mode makes forward detectors such as LHCb and ALICE nearly ideal to probe the negative $x_F$ region, which is essentially uncharted despite its interest.

The \AFTER energy range stands half way between the c.m.s. energies of SPS and the Relativistic Heavy Ion Collider (RHIC), allowing, in particular, for detailed studies of the bottomonium production and dynamics and offering a very large {\it physical} acceptance for detectors covering between 1 to 3 units of rapidity.

In~\cite{Massacrier:2015qba,Kikola:2017hnp}, we have focused on the possibility offered by the LHC nine-months-per-year proton program during which \AFTER would be able to study the production of quarkonia in $pp$, $pd$ and  $pA$ collisions with an unparalleled statistical accuracy down to $x_F \to -1$.  
Such high-precision quarkonium-production measurements in $pp$ are essential~\cite{Andronic:2015wma,Brambilla:2010cs,Lansberg:2006dh} to solve longstanding puzzles in $J/\psi$ and $\Upsilon$ production and for the understanding of their behavior in the (Cold) Nuclear Matter (CNM). 

\AFTER can also provide extremely relevant information on the Quark-Gluon Plasma (QGP), which should be created  at $\sqrt{s_{NN}} =$~72~GeV in Pb$A$ collisions. Despite the decrease in the c.m.s. energy, yields similar to that of LHC at $\sqrt{s_{NN}} =$~2.76 and 5.5~TeV and RHIC at $\sqrt{s_{NN}} =$~200~GeV are to be expected. These are two orders of magnitude larger than the ones obtained at RHIC at $\sqrt{s_{NN}} =$~62~GeV.
The first results from the LHC at $\sqrt{s_{NN}} =$~2.76 and 5.02~TeV~\cite{Andronic:2003zv} confirmed that the pattern of quarkonium (anomalous) suppression at high energy is very intricate with subtle $y$ and $p_T$ dependencies.
Low energy experiments, where the recombination process~\cite{Andronic:2003zv} are not expected to be significant,
can then play a key role in understanding the underlaying physics processes. Such a role will become central if it is possible to carry out  measurements of \chic\ and even \chib\ production and suppression in heavy-ion collisions --2 measurements thus far not performed in any other experimental configuration. The quest for the sequential suppression of quarkonia as a QGP thermometer would then become realistic again.

In addition, the measurement of the Drell-Yan-pair production in asymmetric nucleus-nucleus collisions
provides a unique opportunity to test the factorisation of the initial-state nuclear effect such
as the nuclear PDFs. Such a factorisation is routinely assumed at high energies but a number of non-linear effects can violate it. If these are significant, the nuclear effects from both colliding nucleus do not add up linearly and this would simply prevent us to use the information gained from proton-nucleus or electron-nucleus collisions to characterise the initial stage of nucleus-nucleus collisions. Because of a very high background at the LHC in the collider mode, such measurements are not even feasible in PbPb collisions.
At the LHC, the studies of asymmetric nucleus-nucleus systems are not planned, and such a measurement is very difficult at RHIC. With the lower energy at \AFTER, the studies may be within reach if sufficient different systems are used (\pPb, \Pbp, \PbXe, \pXe, \dots) to pin down and quantify factorisation-violation effects.

The structure of the paper is as follows. In section 2, we introduce the case for heavy-ion physics for \AFTER. In section 3, we discuss our simulation set-up and present performance studies for quarkonium and Drell-Yan probes. We conclude in section 4. 

\section{Quark-gluon-plasma studies with \AFTER}
\label{sec:AA}
\subsection{General considerations}

One of the main incentives to study relativistic heavy-ion collisions is the unique opportunity to probe highly-excited and dense nuclear matter under controlled laboratory conditions. If the temperature and density are high enough,  Quantum ChromoDymanics (QCD), the theory of the strong interaction, predicts the existence of a new phase of matter, the so-called quark-gluon plasma (QGP), in which most of the quarks and gluons, 
normally confined within hadrons, are liberated. Lattice-QCD calculations indeed show that there is a rapid rise of the entropy density when the temperature reaches 160 MeV. Beyond this temperature the effective number of degrees of freedom saturates near the number of quark and gluon helicity and colour states and the entropy density approaches the Stefan-Boltzmann limit. 

Relativistic heavy-ion collisions offer an ideal environment 
to reach the conditions of the phase transition, both in terms of the necessary temperature and volume  for the system thermalisation.
Even at moderately high energies such as the ones obtained at the CERN SPS, the average multiplicity in proton-proton ($pp$) collisions at a c.m.s. energy $E_{c.m.s.}\sim 20$ GeV 
is on the order of 3 per unit rapidity with an average momentum close to~$0.5$~GeV/$c$. This leads to a volume for the system of approximately $1$ fm$^3$ and an energy density $\varepsilon$ on the order of 0.4~GeV.fm$^{-3}$, that is roughly 3 times the density of the normal nuclear matter. 
Heavy nucleus-nucleus ($AA$) collisions, at the same energy per nucleon, rather leads to an energy density of $\varepsilon\sim 2$ GeV.fm$^{-3}$ and an initial volume, before the expansion, on the order of 150 fm$^3$.

Among the expected signatures of the QGP formation, 
quarkonia play an essential role. Since the seminal work of T. Matsui and H. Satz in 1986~\cite{Matsui:1986dk}, the behaviour of quarkonia at high temperature  has been considered a signal for deconfinement in heavy-ion collisions. The original idea behind this proposal is that Debye colour screening in a deconfined plasma would reduce the binding of quarks, and thus also affect the formation of heavy-quark bound states. This effect on quarkonia emerged in the QGP would result in the suppression of the quarkonia yields relative to those in the absence of the plasma formation.

The experimental verification of the Matsui-Satz prediction since the late 1980's certainly is one of the most important quests in particle physics in the last 30 years. Huge progress was achieved since then at the SPS, RHIC and LHC in understanding the properties of matter at extreme conditions~\cite{Braun-Munzinger:2015hba}, in particular that produced in high energy proton-nucleus and nucleus-nucleus collisions. Yet, crucial aspects of the resulting system remain unclear. 
Besides, the mechanism of quarkonium production and suppression in $pp$ \cite{Brambilla:2010cs,Lansberg:2006dh}, $pA$ and $AA$~\cite{Andronic:2015wma} collisions is still far from understood despite significant efforts from the theoretical and experimental sides.

One of the biggest challenge is to find a good baseline to properly use heavy quarkonia to diagnose the QGP. 
This baseline should allow us  to correct the yields for the effects characteristic of heavy-quarkonium production and evolution 
in hadronic matter when QGP is absent. 
Such effects, a priori measurable in proton-nucleus collisions, include 
(i) the modification of the nuclear parton densities, commonly known as shadowing, anti-shadowing, EMC effect or Fermi motion -- depending on the probed kinematics;
(ii) the multiple scattering of partons or of the  heavy-quark pair in the nucleus before or after the hard scattering, which leads 
to an energy loss or the break-up of the formed quarkonium state; 
(iii) the interaction with other particles produced in the collision -- the so-called comovers.
 
The interplay between these effects and the quarkonium suppression in the QGP is still under debate~\cite{Andronic:2015wma}. 
Only high precision data on different quarkonium states, in different $pA$ and $AA$ systems and in different 
kinematics regions ($y$, $P_T$ and $\sqrtsNN$) would clear up these debates.

Another source of complexity indeed results from a possible new charmonium-production mechanism suggested to be at work at high c.m.s. energy. Indeed, $c$ and $\bar{c}$ quarks are so abundantly produced that they could (re)combine into charmonia. In other words, a deconfined nuclear state could lead to more produced $J/\psi$ in some kinematical domains.

\subsection{The case for \AFTER}

In this context, it is important to keep in mind the central role played by fixed-target experiments 
in this field.  In general, they provide very high precision measurements whose relevance is clear from the above discussion. 
The series of experiments NA38, NA50 and NA60 at the CERN SPS program have indeed shown the first tangible 
signs of anomalous charmonium suppression in heavy-ion collisions~\cite{Abreu:1997jh}. Yet, as of today, this anomalous character could not unambiguously be related to a specific phenomenon or ensemble of phenomena.

Without any surprise, \AFTER, as it happened with these SPS experiments, present significant assets with respect to the present collider experiments, LHC and RHIC. Its c.m.s. energy of 72 GeV allows one to study quarkonium observables in the region of a possible phase transition where charmonium recombination is not expected to play an important role. High luminosities, inherent to the fixed-target mode, are not only necessary to increase the statistical precision but also to provide us with a better control of acceptances and efficiencies. 
The NA60 experiment together with experiments at HERA and FNAL provide ample proof to that. 

Moreover, the NA61 experiment at SPS  energies and the RHIC beam-energy-scan program are the natural complement of \AFTER. 
Together they can provide a much awaited detailed picture of the phase transition region from SPS to RHIC 
energies and clarify its nature.

Taking into consideration the \AFTER c.m.s. energy and luminosity, the first inclusive high-precision  program for quarkonia is at reach. It is also complemented by heavy- and light-flavour studies as we detail now.

\subsubsection{The \AFTER quarkonium-physics case}

\paragraph{Quarkonium excited states.}

The scope of \AFTER for quarkonium excited states is fully inclusive and encompasses not only the study of \jpsi, \psip and $\ups(nS)$ states but also the  $\chi_{c,b}$ states and the corresponding associated-production channels whose study requires high luminosities, good detector performances (granularity, resolution, PID), wide acceptance and reliable efficiency controls.

It is a known fact that the $J/\psi$ and $\Upsilon(nS)$ states receive significant feed-down contributions. 
The observed ground states \jpsi and $\ups(1S)$ are in 
both cases significantly produced through feed down from higher excited states. This is also true for
$\ups(2S)$ and $\ups(3S)$.  Understanding quarkonium formation and suppression will most probably only be achieved once the
excited-state effect are properly accounted for. This begins with careful studies 
of the $\psi'$, $\ups(2S)$ and $\ups(3S)$ yields which is clearly possible with \AFTER.

In section \ref{sec:sim}, we present our projection for the study of $\psi(nS)$ and $\Upsilon(nS)$ states
in $AA$ collisions at \AFTER energies. These complement a previous study~\cite{Massacrier:2015qba} for $pp$ and $pA$ collisions.

Yet, a full description of the mechanisms for the formation of the quarkonium $S$ states cannot be dissociated from that of $P$ states. Direct measurements are probably possible with \AFTER with a reduced combinatorial background thanks to the reduced energies compared to the LHC or RHIC. 
The study of topological properties of the decay distribution such as polarisation~\cite{Faccioli:2012kp,Faccioli:2008dx,Faccioli:2014cqa} discussed below can complement such direct studies without requiring to detect photons.

\paragraph{Quarkonium polarisation.}

Quarkonium polarisation -- despite being considered for a long time to be a smoking-gun signal --  has so far eluded a thorough explanation despite decades of experimental and theoretical efforts. In the recent years, it was realised that both polar and azimuthal distributions 
should be measured altogether \cite{Faccioli:2008dx}. 
\AFTER, being a high precision experiment, can play in this respect a very important role in clarifying the evolution of the angular anisotropy parameters for the decay of heavy quarkonia from low to high energies and thus enlighten the production mechanism for these states.

Moreover, as mentioned above, the $\chi_c$ and $\chi_b$ production in nucleus-nucleus collisions necessarily impacts the interpretation of the $J/\psi$ and $\Upsilon$ suppression as a probe of QGP formation. The observation of the $\chi_c$ and $\chi_b$ suppression patterns in nucleus-nucleus collisions can help to confirm or falsify the sequential quarkonium melting scenario and, therefore, discriminate between the QGP interpretation and other options. 

In addition to a direct observation of the $\chi_c$ and $\chi_b$ signals in their radiative decays to $J/\psi$ and $\Upsilon(nS)$ which is admittedly challenging in heavy-ion collisions,  it may be possible to determine the relative yield of $P$ and $S$ states by only performing  dilepton polarisation measurements.  The predicted differences in the nuclear dissociation patterns of $S$ and $P$ states -- owing to their different binding energies-- should result in a change of the observed $J/\psi$ and $\Upsilon$ polarisations from proton-proton to central nucleus-nucleus collisions. This may provide an indication for quarkonium sequential suppression in the QGP. For such observations to be conclusive, the impact of other QGP- and nucleus-related phenomena on the $J/\psi$ and $\Upsilon$ polarisation should carefully be investigated alongside
in $pA$ collisions or on the states receiving few or no feed down, like the $\psi'$ for instance.

\subsubsection{Open charm and beauty production}

The study of $D$ mesons is a natural continuation of the study of \jpsi and \psip formation and dissociation. 
$D$ mesons has been proposed~\cite{Satz:2005hx} as a baseline to understand QGP effects on quarkonium. 
In particular, it was proposed to measure their yield in $pA$ and $AA$ collisions versus rapidity to get some insights
on CNM affecting heavy-quark production, which can be common to quarkonium production. Such studies can be 
complemented with the measurements of azimuthal anisotropies as a function of rapidity to constraint transport properties of the QGP such as the shear viscosity and the heavy-quark diffusion coefficients. The measurement of the nuclear-modification factor, \RAA\, for open-heavy flavour hadrons can also provide insights into the nature of the heavy-quark interaction with the surrounding matter. The data collected at RHIC at the energy of \sqrtsNN = 200~GeV~\cite{Adamczyk:2014uip,Adare:2006nq} show a strong suppression at transverse momenta larger than 3 GeV/c. These results can be considered as the  evidence of heavy-energy loss in the hot and dense QGP. However, the details of the interactions with the QGP are not well understood; the major difficulty is to determine the role of the gluon radiation (radiative energy loss) and of the collisional energy loss (due to collisions with other objects in the QGP). A few different models with distinctive assumptions describe reasonably well the current data~\cite{Adamczyk:2014uip}. New observables --preferably for open charm and beauty separately-- are thus required to constrain these theoretical approaches. 

\AFTER\ will be capable of delivering high-quality data to address this issue. The measurements of nuclear modification factors for D mesons as a function of transverse momentum and rapidity with \AFTER\ will provide extremely valuable inputs since the radiative and collisional energy loss are expected to exhibit different \pT\ and $y$ dependence~\cite{Kikola:2015lka}. These high statistics data will also facilitate studies of heavy-flavour azimuthal correlations ($D-D, J/\psi - D$) which will give another handle on heavy-quark in-medium interactions. These will indeed allow for precise determinations of the transport properties of the QGP, including the energy-loss transport coefficient $\hat{q}$ and the charm-quark-diffusion coefficients. Finally, the measurements of the nuclear-modification factor for $b$-hadrons in \pA\ interactions will be measurable with a good precision~\cite{Kikola:2015lka,Massacrier:2015qba}. $b$-hadron \RAA\ measurements	 via non-prompt $J/\psi$ in \PbA\ collisions will also be at reach with a slightly reduced statistics compared to the prompt ones but with a smaller background. 

\subsubsection{Drell-Yan}
The hadro-production of a large invariant-mass lepton pair, known as the Drell-Yan (DY) process,  is a clean, precise and controllable probe of the short-distance dynamics and the partonic structure of hadrons. In particular, the DY process on nuclear targets is an ideal tool to quantify the initial-state effects taking place during such proton-nucleus reactions. Let us cite the nuclear modification of the parton densities or coherence effects. Indeed, the production of the lepton pair by an electroweak gauge boson does not suffer from final-state interactions, typically associated with the phenomena of energy loss or absorption .

Beside being a typical background for the study of quarkonia, it is also an important process \emph{per se} where the factorisation of initial-state nuclear effects can be tested. With a nuclear target, the \AFTER kinematic with a LHCb-like detector cover $2 \lesssim y_{\rm lab} \lesssim 5$ correspond to the antishadowing region for the parton in the targer ($x_2\in [0.05:0.2]$) for invariant masses slightly above the charmonium family. Increasing the dilepton mass allows one 
to probe the EMC region, where the nuclear modification of the PDFs was historically observed for the first time. The use of \Pb\ beam allows one to probe a region where shadowing effects are expected. Overall, the study of \pA, \Pbp and \PbA collisions altogether will offer an unique playground to test the factorisation of nPDF effects.

\subsubsection{Soft-probe studies}

The aforementioned \AFTER studies on hard probes will be complemented with that of 
soft probes in heavy-ion collisions. In fact, the bulk of low-momentum particles produced in \AA\ collisions at RHIC and at the LHC is supposed to originate from a fluid-like medium, whose transverse expansion is largely driven by the density gradients produced in the earliest stage of the collision \cite{Ollitrault:1992bk,Kolb:2000sd}. The harmonic analysis of particle production in \AA\ collisions typically focuses  on the Fourier decomposition of the azimuthal angle $\phi$ dependence over a narrow region in pseudorapidity around $\eta=0$. The $v_n$ coefficients of this decomposition play an important role in understanding the collective behavior of the system. The study of both even and odd order coefficients~\cite{Naselsky:2012nw} provide tight constraints on the collective dynamics and should be determined for that reason.

\begin{figure}[ht!]
\centering
\includegraphics[width=0.75\textwidth]{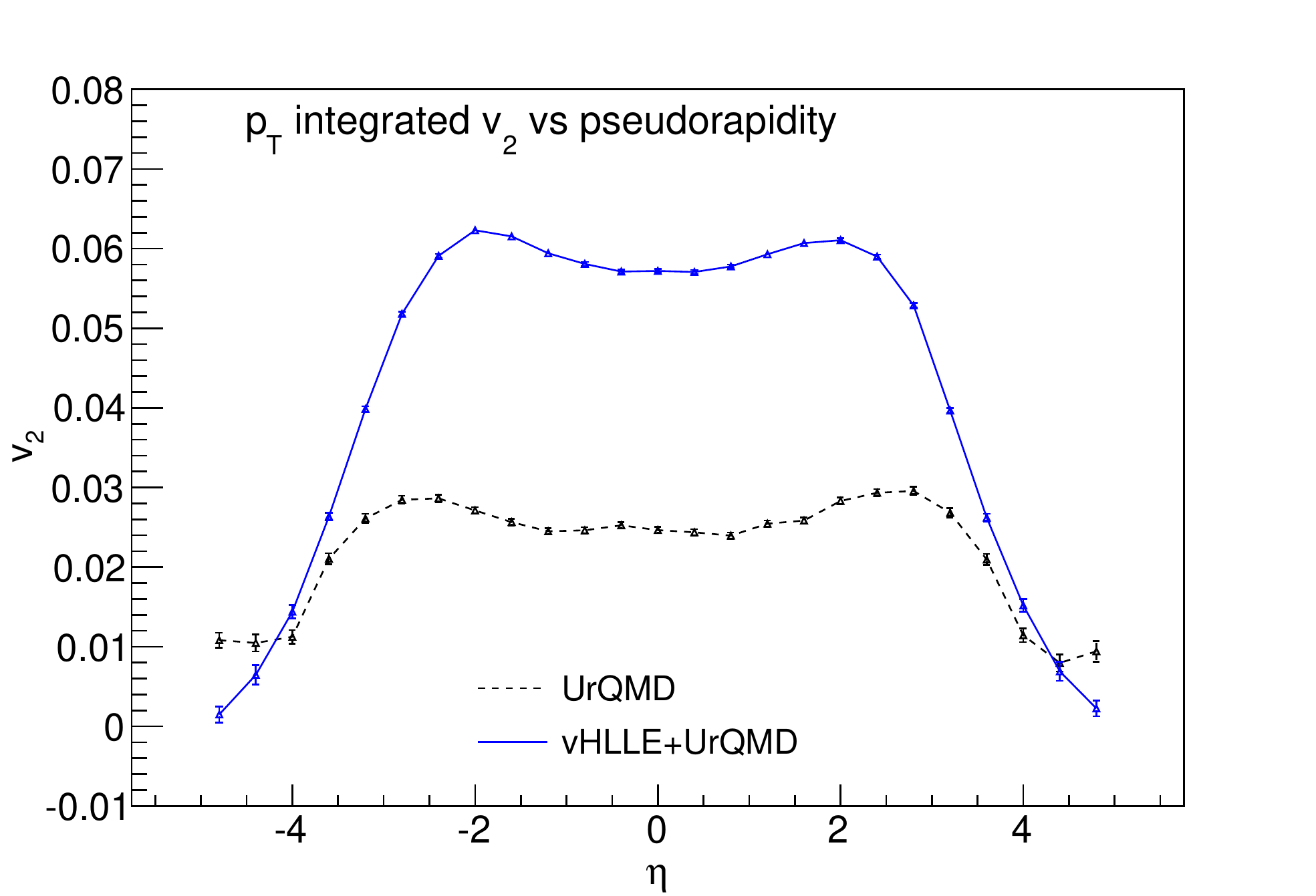}
\caption{$p_{T}$-integrated ($p_{T}$ range=[0.2, 2.0] GeV) elliptic flow (calculated using event plane method) of all charged hadrons as a function of c.m.s. pseudorapidity in \PbPb collisions at $\sqrt{s_{NN}}$=72 GeV and centrality 20-30\% (which in the model correspond to impact parameter range b=[6.8, 8.3] fm). The dashed curve corresponds to a pure cascade case (\textit{UrQMD}) and is compared to the case (solid blue curve) with an intermediate hydrodynamic phase of the evolution  \textit{vHLLE+UrQMD} (hydro+cascade model)~\cite{Karpenko:2015xea}.}
\label{fig:v2HydroPrediction}
\end{figure}

This requires, in particular, a baseline comparison of $v_2$ (and $v_n$) to existing models in $AB$ collisions, for two otherwise opposite assumptions - with and without a hydrodynamic phase of the evolution.
Figure~\ref{fig:v2HydroPrediction} shows predictions of charged hadrons $v_2$ versus pseudorapidity in semi-central \PbPb\ collisions at $\sqrt{s_{NN}}=72$~GeV for a pure cascade (\textit{UrQMD}) case and for the case when a hydrodynamic evolution is included (\textit{vHLLE+UrQMD})~\cite{Karpenko:2015xea}.
\AFTER will be able to continue the existing studies on $v_n$ at energies between the RHIC Beam Energy Scan (\textit{BES}) and the top RHIC and LHC energies not only with high precision but also in a broad rapidity range. Doing so, it will provide a detailed account of the collective evolution between the two energy ranges, around the expected transition line in the QCD phase diagram, between the existing high precision SPS results and the LHC. 

\subsubsection{Going further}

Other points that will be envisaged include:
  
During the ramp up in energy of the LHC, \AFTER can perform an energy scan to search for a possible critical point along the phase transition line, along the same lines as the RHIC BES program. Even if this scan may be an experimental challenge owing to the possible beam instabilities during this phase, it can provide relatively high statistics allowing for direct searches of the critical point and to look for phase-transition signatures. As for the soft probes study at 72 GeV, it would be complementary to the RHIC BES II program.

A study of the chiral symmetry restoration \emph{\`a la} NA60 can be performed in \AFTER by studying the dimuon mass spectrum in $AA$ for the $\rho$, $\omega$, $\phi$ systems. Different measurements can be performed, namely the change in the branching ratio between leptonic and hadronic channels and the
change in the width of the mass peaks.  This requires significant statistics, a high resolution and a careful control of the background processes including the charm contribution. A comparison with the STAR observation at RHIC (excess with 2$\sigma$ at $\sqrtsNN=62$~GeV -- without charm background subtraction) would be extremely useful. Since the NA60 results~\cite{Arnaldi:2006jq} were obtained with a resolution of 20 MeV at the $\omega$ and $\phi$ peaks and \AFTER studies should aim at such precision to be competitive. The particle identification which could be done with a LHCb-like detector would also considerably simplify the analysis and overcome many of the drawbacks present in NA60.

Overall, \AFTER can considerably enrich the existing studies devoted to the search for signatures of the chiral symmetry restoration which plays a fundamental role in explaining the origin of 95\% of the meson and baryon masses. Aided by lattice studies and a considerable wealth of information collected since decades on the fundamental aspects of chiral symmetry in the strong sector~\cite{Borasoy:2007yi}, \AFTER is in a unique position to clarify the role of the chiral-symmetry restoration in the dynamics of meson and baryon interactions and also its importance in the interplay between chiral-symmetry restoration and deconfinement transition~\cite{Karsch:1994hm}.
 
Finally,  \AFTER, due to its extended kinematic range can scan a much larger chemical baryon potential range in the QGP phase diagram compared to measurements at mid-rapidity which are presently available in most of the experimental results.

\section{Simulations}
\label{sec:sim}
In this section, we present feasibility studies for quarkonium and Drell-Yan production via the di-muon decay channels in heavy-ion collisions with a 2.76 TeV per nucleon LHC lead beam on a fixed target ($\sqrt{s_{NN}} =$~72~GeV). 
In Ref.~\cite{Massacrier:2015qba}, we have presented simulations for \pp collisions at $\sqrt{s}=115$ GeV resulting from 7 TeV protons impinging on a hydrogen target with a LHCb-like detector. In the following, we follow a similar procedure which we have updated for the heavy-ion case as described below. In addition, we elaborate a little on the possibilities offered by the ALICE detectors used in the fixed-target mode.

\subsection{Simulation framework}

The simulations presented in this paper are performed for \pp collisions at $\sqrt{s_{NN}} =$~72~GeV. Projections for \pA and \AA collisions at $\sqrt{s_{NN}} =$~72~GeV are done by applying a nuclear scaling factor to the \pp simulations, assuming neither nuclear nor isospin effects. 

Different di-muon sources are separately studied in order to keep a good control over the input distributions and the normalization of these different sources. As a signal we consider either quarkonia or Drell-Yan (which is a background in the case of quarkonium studies). The background under the signal is twofold: the correlated background -- charm and beauty pair production ($c$+$\bar{c} \rightarrow D^++D^- \rightarrow l^+l^-$ and $b$+$\bar{b} \rightarrow B^++B^- \rightarrow l^+l^-$), and the combinatorial background, mainly muons from pion and kaon decays. 

The quarkonium signals as well as the Drell-Yan and charm-- and beauty--pair production (\cc and \bb) are simulated with {\sc\small HELAC-Onia}~\cite{Shao:2012iz,Shao:2015vga} which provides Les Houches Event Files~\cite{Alwall:2006yp} as output. These events are then processed with {\sc\small Pythia} 8~\cite{Sjostrand:2007gs} to perform the hadronisation, to account for initial/final-state radiations and to decay the resonances. On this step additional muons from the underlying {\sc\small Pythia} event can be produced. We have checked that the contribution of combinations of these muons with muons from initial quarkonia or \cc, \bb pairs is negligibly small and thus is not included in the correlated background.

\jpsi, \psip, \ups states and the \cc continuum rates are obtained in a data-driven way as in~\cite{Massacrier:2015qba}. Quarkonia processed in {\sc\small Pythia} 8 are forced to decay into the dimuon-decay channel and the simulated yields are then weighted by the corresponding cross section multiplied by the Branching Ratio (BR)~\cite{Beringer:1900zz}. The open--beauty simulation is performed with a Leading Order (LO) matrix element normalised to a Next-To-Leading-Order (NLO) $K$ factor found to be 1.83~\cite{Massacrier:2015qba}.
The Drell-Yan simulation is performed with the process $q\bar{q} \rightarrow \gamma^{\star}/Z \rightarrow \mu^{+} \mu^{-}$ at LO, with the CTEQ6L1 pdf set. In this case, the $K$ factor is fond to be 1.2. In order to decrease the time for the simulation in the di-muon invariant mass ($M_{\mu ^{+} \mu ^{-}}$) region of interest, the simulation is done with a $M_{\mu ^{+} \mu ^{-}} >$~1.5~GeV/$c^{2}$ requirement. 

The combinatorial background is obtained from minimum bias \pp collisions generated with {\sc\small Pythia} 8, using the process \texttt{SoftQCD:nonDiffractive} with the MRSTMCal.LHgrid LHAPDF (6.1.4) set~\cite{Buckley:2014ana}. The dominant source of combinatorial opposite-sign di-muon pairs are $\mu^{+/-}$ coming from $\pi^{+/-}$ or $K^{+/-}$ decays.

The {\sc\small Pythia} 8 output is then processed via a fast simulation framework. This last step is performed to account for realistic detector-resolution and particle-identification performances. The detector response -- momentum resolution, $\mu$ identification efficiency and $\pi/K$ misidentification probability with $\mu$ -- is simulated with a detector setup similar to the LHCb detector~\cite{Alves:2008zz} with a pseudorapidity coverage of 2~$< \eta_{\rm lab} <$~5. The minimum transverse momentum of single muons is required to be greater than 0.7~GeV/$c$. 
The considered momentum resolution is taken as $\delta p / p =$ 0.5 \%, the LHCb reporting: $\delta p /p \sim$~0.4~(0.6)\% for a momentum of 3~(100)~GeV/$c$~\cite{Archilli:2013npa}. The considered single $\mu$ identification efficiency of $\epsilon _{\mu^{+/-}} =$~98\% is an average LHCb efficiency for muons coming from \jpsi decays, for $p > 3$~GeV/$c$ and $p_\mathrm{T} > 0.8$~GeV/$c$~\cite{Archilli:2013npa}.
In the case of muons from hadronic decays, the muon is not considered in the simulation if a $\pi$/$K$ decays before 12~m. This corresponds to a decay before the LHCb calorimeter which can be rejected by the LHCb tracking system. If the $\mu$ is produced beyond 12~m or if a $\pi/K$ is misidentified with $\mu$ in the muon stations, the following momentum-dependent $\pi/K$ misidentification probabilities~\cite{Aaij:2014jba} are applied: $P_{MID}(\pi \rightarrow \mu)(p) = (0.5 + 6.63 \exp(-0.13p)) \%$ and $P_{MID}(K \rightarrow \mu)(p) = (0.5 + 8.6 \exp(-0.11p)) \%$, for $\pi$ and K respectively. The di-muon efficiency is taken as: $\epsilon _{\mu^{+}\mu^{-}} = \epsilon_{\mu^{+}} \times \epsilon_{\mu^{-}}$.
It is assumed in these simulations that the detector performance does not decrease with the event multiplicity.

\begin{table*}[!hbtp] 
\caption{Total cross section for different processes in \pp collisions at $\sqrt{s}$~=~72~GeV obtained from {\sc\small HELAC-Onia} simulations, and  in the case of the minimum bias simulation from {\sc\small Pythia} 8.} 
\label{tab:crosssection}
\center{\renewcommand{\arraystretch}{1.5}
\begin{tabular}{c|p{4.cm}}\footnotesize
         & $\sigma_{tot}$ (mb) \\ \hline \hline
\jpsi & 5.51 $\times$ 10$^{-4}$\\ 
\psip & 5.90 $\times$ 10$^{-5}$ \\
$\Upsilon$ (1S) & 6.87 $\times$ 10$^{-7}$ \\
$\Upsilon$ (2S) & 1.82 $\times$ 10$^{-7}$  \\
$\Upsilon$ (3S) & 7.56 $\times$ 10$^{-8}$\\
Drell-Yan (M $>$ 1.5 GeV/c$^{2}$) & 6.66 $\times$ 10$^{-6}$\\
$c\bar{c}$ &  1.07 $\times$ 10$^{-1}$  \\	
$b\bar{b}$ &  9.44 $\times$ 10$^{-5}$ ($gg \rightarrow b\bar{b}$) \newline 5.86 $\times$ 10$^{-5}$ ($q\bar{q} \rightarrow b\bar{b}$) \\
minimum bias &  25.38\\ 
\end{tabular}}
\end{table*}

Projections for \pA and \AA collisions are done by applying a nuclear scaling factor, accounting for the number of binary collisions, to the cross sections obtained from the \pp simulations. 
Then, the signal and background sources are normalised to the desire integrated luminosity according to the production cross section of the process (taking into account the initial phase space cuts, if any). The values of the cross sections that are obtained from the \pp simulations at $\sqrt{s_{NN}} =$~72~GeV are reported in \ct{tab:crosssection}, they are integrated over rapidity and $p_{T}$. In the case of quarkonium, Drell-Yan, open charm and beauty simulations, the nuclear scaling factor for minimum-bias \pA collisions is $A$ and for \AB collisions it is $A B$, where $A$ and $B$ are the beam and projectile mass numbers, respectively. These scaling factors correspond to the absence of hot/cold nuclear and isospin effects.
The scaling factor for the combinatorial background is $A \times N_{coll}^{pA}$ and $AB \times N_{coll}^{AB}$ for \pA and \AB collisions, respectively, where $N_{coll}$ is an average number of binary collisions obtained from Glauber simulations~\cite{Alver:2008aq,Loizides:2014vua,Kikola:2015lka} for minimum bias collisions for a given colliding system.

In what follows, we will study the case of Xe gas target, one of the heavier noble gases. The LHC heavy-ion runs are expected to last for about a month per year, which we take as $10^{6}$~s. In order to provide a baseline at the same energy, 2.76~TeV proton beam is delivered for about a week per year, and our estimated running time for proton-beam collisions on a hydrogen gas target is $0.25 \cdot 10^{6}$~s. The expected instantaneous and yearly (as defined before) luminosities for \pp and \PbXe collisions at $\sqrt{s} =$~72 GeV are gathered in Table~\ref{tab:lum}. A month of running with the LHC Pb beam on a gas target provides an access to heavy-ion studies with a yearly integrated luminosity as high as 30~nb$^{-1}$ with an integrated luminosity of the \pp baseline reaching 250~pb$^{-1}$.
Furthermore, studies of \pA collisions at the same centre-of-mass energy of $\sqrt{s} =$~72 GeV would be of particular interest. 
The instantaneous luminosity for \pA collisions (with a Xe, Kr, Ar or Ne gas target) is included in Table~\ref{tab:lum}. These instantaneous luminosities are expected to be achieved with a 1-m-long HERMES-like internal gas target~\cite{Barschel:2015mka}. Such an open-storage-cell system with a dedicated pumping system offers a possibility to reach a higher gas pressure\footnote{In~\cite{Massacrier:2015qba}, a SMOG-like system was considered and resulted in lower instantaneous luminosities}.
Given these high instantaneous luminosities, one can obtain the same yields of \jpsi, \psip, $\Upsilon$ and Drell-Yan in \pA collisions as in the \pp case with relatively short runs. 
For example, in the case of the \pXe system considered in this paper, a running time of 13 hours is enough, which corresponds to an integrated luminosity of $\sim$ 2~${\rm pb}^{-1}$. For Kr, Ar and Ne cases that would be 20 hours, 42 hours and 84 hours of running, respectively. These running times correspond to integrated luminosities of 3~${\rm pb}^{-1}$, 6.5~${\rm pb}^{-1}$ and 150~${\rm pb}^{-1}$ for \pKr, \pAr and \pNe collisions, respectively \footnote{The mentioned instantaneous luminosities correspond to an inelastic rate of 400 MHz. This may however be limited by detector capabilities. In such case, longer running times may be needed in order to achieve the quoted integrated luminosities for \pA systems.}. 
Table~\ref{tab:lum} also includes average number of binary collisions for minimum-bias events, for each of the considered colliding systems: \pp, \pXe and \PbXe. 

The projections which are presented here for \pXe collisions give the highest combinatorial background out of the mentioned \pXe, \pKr, \pAr, \pNe cases with the 2.76~TeV proton beam, due to the highest number of binary collisions. 
The important advantage of the fixed-target mode is the possibility to change relatively easily from one target to another. This would allow us to study different systems, such as \PbA, \Pbp and \pA collisions at the same energy, in different years of LHC running.

\begin{table*}[!hbtp] 
\caption{Expected instantaneous and yearly (as described in the text) luminosities for \PbA, \pA and \pp collisions with 2.76~TeV beams, and an average number of binary collisions for \PbXe and \pXe minimum bias collisions from the Glauber Monte Carlo calculations.} 
\label{tab:lum}
\center{\renewcommand{\arraystretch}{1.5}
\begin{tabular}{c|c|c|c|c}\footnotesize
Beam & Target gas &  $\mathcal L$ $[\rm{cm}^{-2}\rm{s}^{-1}]$  & $\int dt\,  \mathcal L$ & $\langle N_{\rm coll} \rangle$ \\ \hline \hline
\multirow{2}{*} {Pb} & Xe & \multirow{2}{*} {$3 \cdot 10^{28}$}  & \multirow{2}{*} {30 $\rm{nb}^{-1}$} & 165.1 \\
& (Kr, Ar, Ne) &   &  &  \\ \hline
\multirow{2}{*} {$p$} & Xe &  \multirow{2}{*} {$2.34 \cdot 10^{32}$} & 2 $\rm{pb}^{-1}$ & 3.7 \\
 & Kr, Ar, Ne &   & 3 $\rm{pb}^{-1}$, 6.5 $\rm{pb}^{-1}$, 150 $\rm{pb}^{-1}$  &  \\ \hline
$p$ & H & $0.92 \cdot 10^{33}$  & 250 $\rm{pb}^{-1}$ & 1 \\
\end{tabular}}
\end{table*}

\subsection{Quarkonium studies at $\sqrt{s_{NN}} =$ 72~GeV}

For the quarkonium \PbXe simulations where the full background is considered, the combinatorial one is subtracted assuming the like-sign technique. The background can be reconstructed as an arithmetic or geometric mean of all $\mu^+ \mu^+$ and $\mu^- \mu^-$ pairs produced in an event.
That is the worst-case scenario in terms of the expected statistical uncertainties from the combinatorial-background subtraction: the statistical uncertainties of the background scale as the background yield under the signal ($N_{bkg}$) -- $\sqrt{N_{bkg}}$ -- and add to that of the signal when subtracted.  
However, this is the most straightforward method of extracting the combinatorial background in an experiment. It does not require an additional normalisation factor and thus does not suffer from additional systematic uncertainties, as the event-mixing technique for example.

\begin{figure}[ht!]
\centering
\includegraphics[width=0.65\textwidth]{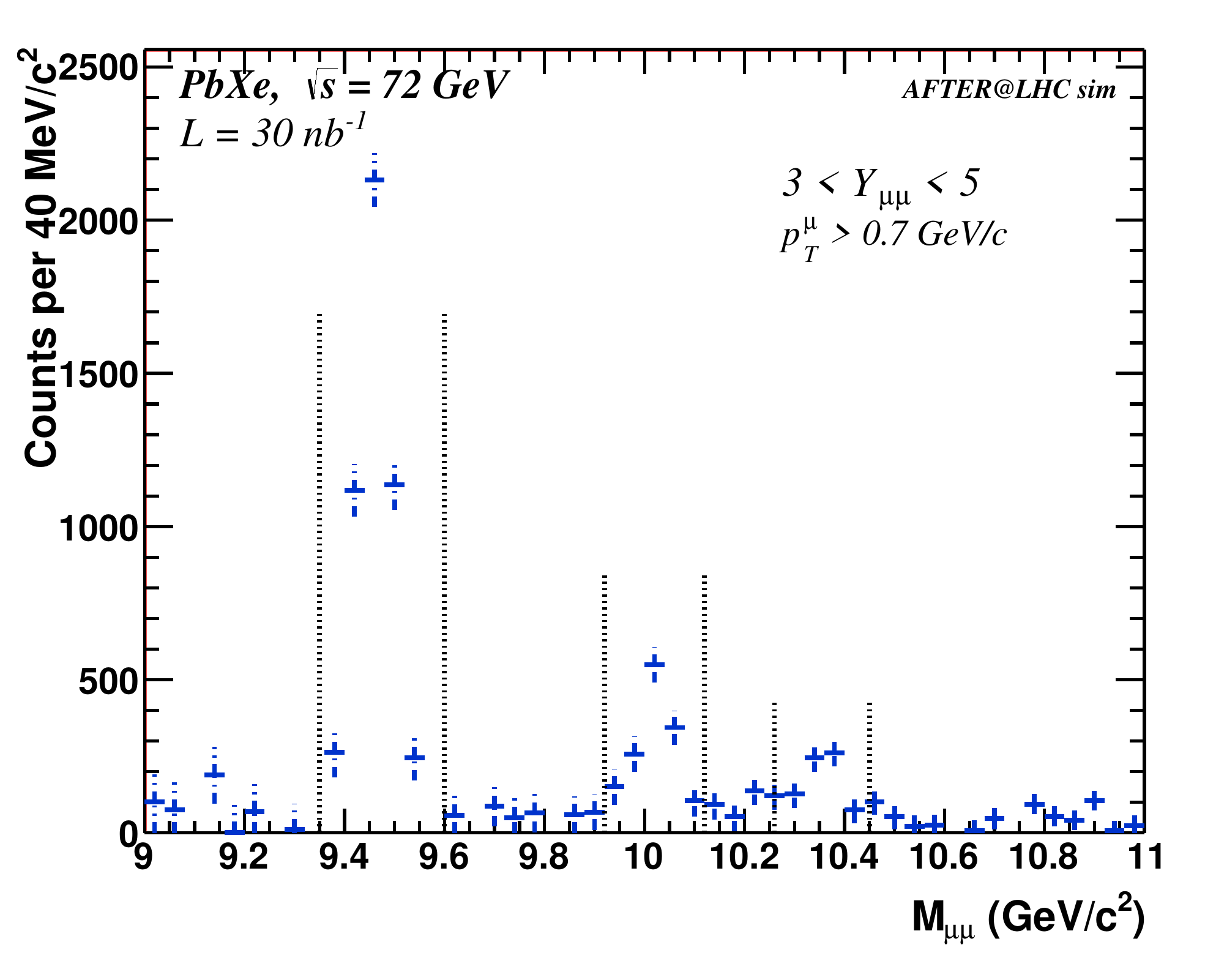}
\caption{$\Upsilon(nS)$ signal after the (like-sign) combinatorial-background subtraction with the expected statistical uncertainties in \PbXe collisions at $\sqrt{s_{NN}} =$ 72 GeV in 3 $< y_{\rm lab} < $ 5 for $\int \mathcal L_{\rm PbXe}$ = 30 ${\rm nb}^{-1}$, assuming $R_{AA}=$ 1.}
\label{fig:Upsilon_HI}
\end{figure}

\begin{figure}[htp]
\centering
\begin{tabular}{ll}
  \includegraphics[width=0.48\textwidth]{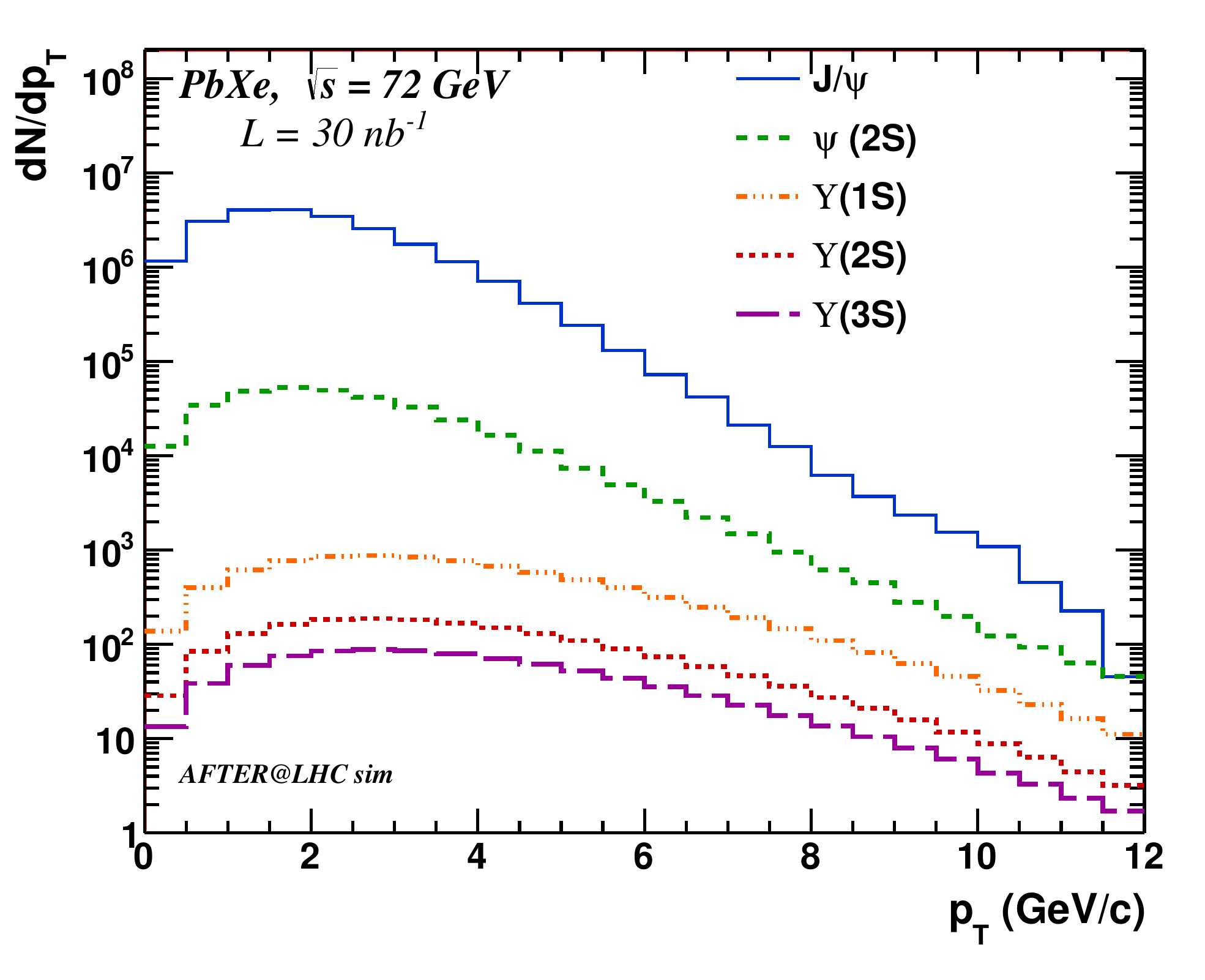}
  \includegraphics[width=0.48\textwidth]{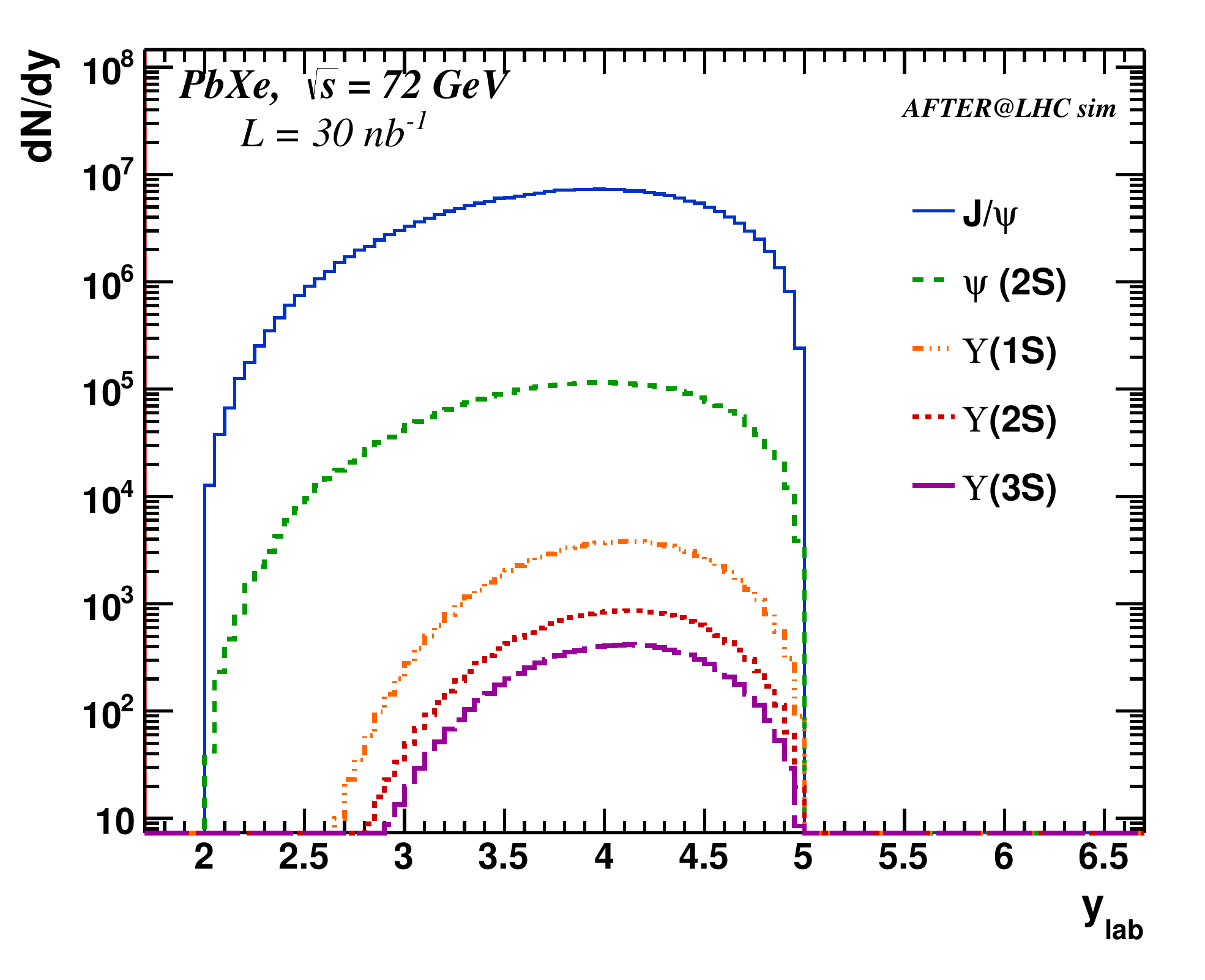}
\end{tabular}
\caption{\jpsi, \psip, $\Upsilon(1S)$, $\Upsilon(2S)$ and $\Upsilon(3S)$ $p_{T}$, $dN/dp_{T}$, (left) and rapidity, $dN/dy_{\rm lab}$, (right) spectra in \PbXe collisions at $\sqrt{s_{NN}} =$ 72 GeV for $\int \mathcal L_{\rm PbXe}$ = 30 ${\rm nb}^{-1}$, assuming $R_{AA}=$ 1.}
\label{fig:quarkonia_kin}
\end{figure}

As can be seen on Figure~\ref{fig:Upsilon_HI}, high $\Upsilon (nS)$ yields are expected in \PbXe heavy-ion collisions with an integrated luminosity of 30 ${\rm nb}^{-1}$. The plot shows the di-muon invariant mass distribution in the $\Upsilon (nS)$ mass range after the combinatorial background subtraction with a like-sign technique assuming $R_{AA}=$ 1. Thanks to the large $\Upsilon (nS)$ yields and large enough signal / background ratios ($S/B$), each of the $\Upsilon$ state -- $\Upsilon (1S)$, $\Upsilon (2S)$ and $\Upsilon (3S)$ -- can easily be reconstructed. The dashed lines represent expected width of $\Upsilon (nS)$ states assuming the LHCb-like single muon $p_{T}$ resolution: the $\Upsilon$ states can clearly be separated. The other background sources, namely correlated $b \bar{b}$ and Drell-Yan pairs, are negligibly small in the $\Upsilon (nS)$ invariant-mass range. The expected yields for $\Upsilon (1S)$, $\Upsilon (2S)$ and $\Upsilon (3S)$ and $S/B$ ratios are summarised in Table~\ref{tab:Upsilon_yields} for \pp, \pXe and \PbXe collisions. They are integrated over the $\Upsilon$ $p_{T}$ and a rapidity range of 3 $< y_{\rm lab} < $ 5, accessible for the $\Upsilon$ production. The $p_{T}$ and rapidity spectra, for each $\Upsilon$ state are shown in Figure~\ref{fig:quarkonia_kin}.

\begin{table}[!htb]
\caption{$\Upsilon(nS)$ yields and $\Upsilon(nS)$ signal over the combinatorial background ratios ($S/B$) for \pp, \pXe and \PbXe collisions at $\sqrt{s_{NN}} =$ 72 GeV and 3 $< y < $ 5, $\int \mathcal L_{pp}$ = 250 ${\rm pb}^{-1}$, $\int \mathcal L_{p{\rm Xe}}$ = 2 ${\rm pb}^{-1}$, $\int \mathcal L_{\rm PbXe}$ = 30 ${\rm nb}^{-1}$.}
\label{tab:Upsilon_yields}
\begin{center}{\renewcommand{\arraystretch}{1.5}
\begin{tabular}{c|c|c|c}
 \multicolumn{1}{c}{Yields} &  & signal & S/B \\ \hline \hline
 \multirow{3}{*}{$\Upsilon(1S)$} & \pp & 1.33 $\times 10^{3}$ & 29.0 \\ 
 & \pXe & 1.39 $\times 10^{3}$ & 7.8 \\
 & \PbXe & 4.33 $\times 10^{3}$ & 1.8 $\times 10^{-1}$ \\ \hline
 \multirow{3}{*}{$\Upsilon(2S)$} & \pp & 2.92 $\times 10^{2}$ & 8.2 \\ 
 & \pXe & 3.06 $\times 10^{2}$ & 2.2 \\
 & \PbXe & 9.56 $\times 10^{2}$ & 5.0 $\times 10^{-2}$ \\ \hline
 \multirow{3}{*}{$\Upsilon(3S)$} & \pp & 1.37 $\times 10^{2}$ & 10.3 \\ 
 & \pXe & 1.44 $\times 10^{2}$ & 2.8 \\
 & \PbXe & 4.49 $\times 10^{2}$ & 6.2 $\times 10^{-2}$ \\ 
\end{tabular}}
\end{center}
\end{table}

With the integrated luminosity of 30 ${\rm nb}^{-1}$, extremely large \jpsi and \psip yields are expected. The \jpsi and \psip $p_{T}$, $dN/dp_{T}$, and rapidity, $dN/dy_{\rm lab}$, spectra are shown in Figure~\ref{fig:quarkonia_kin}, assuming $R_{AA}=$ 1. \jpsi and \psip signals can be studied with a good precision over a wide transverse momentum range, up to $\sim$ 12 GeV/$c$, and in the rapidity range limited only by the detector acceptance.

\begin{figure}[htp]
\centering
\begin{tabular}{ll}
  \includegraphics[width=0.48\textwidth]{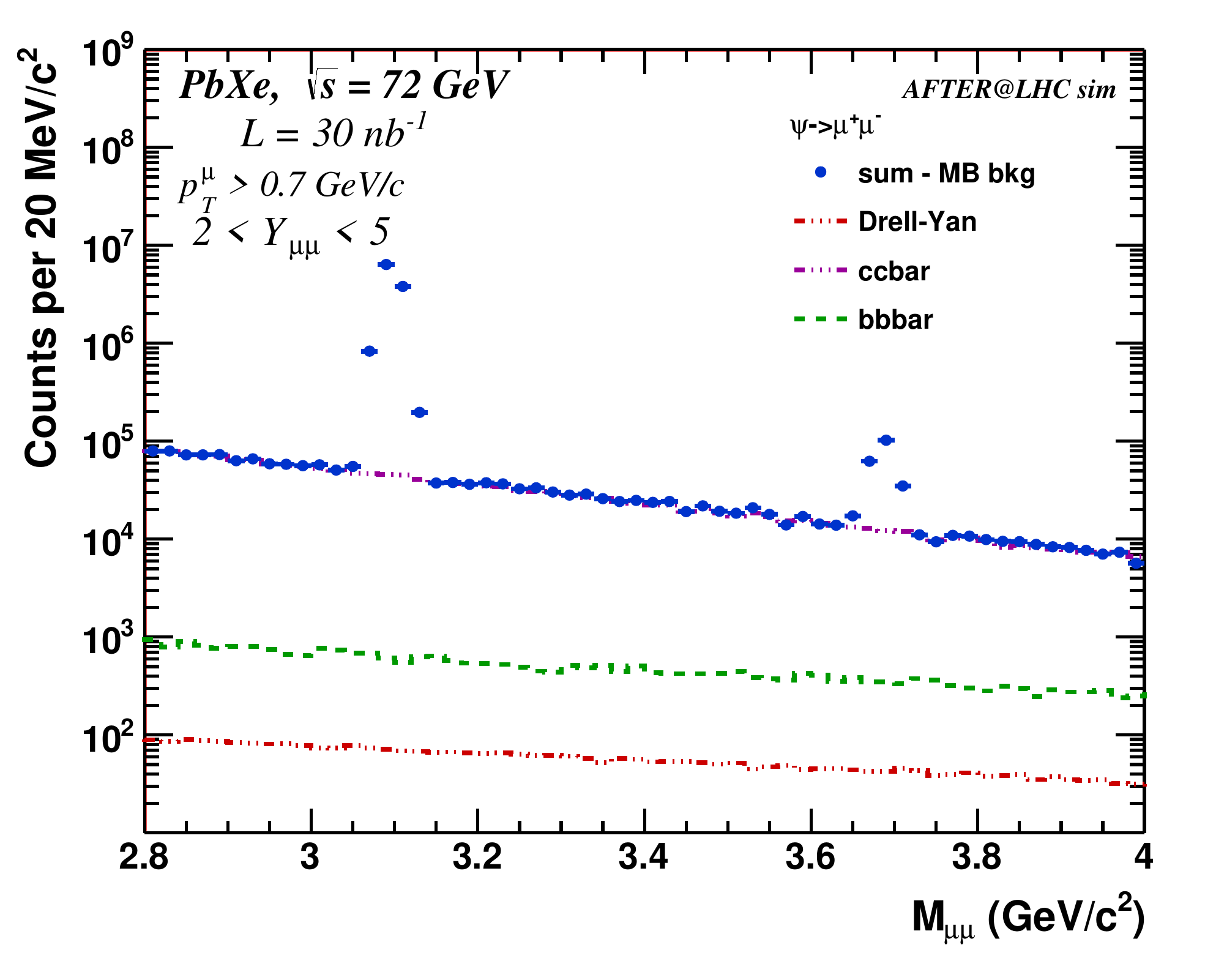}
  \includegraphics[width=0.48\textwidth]{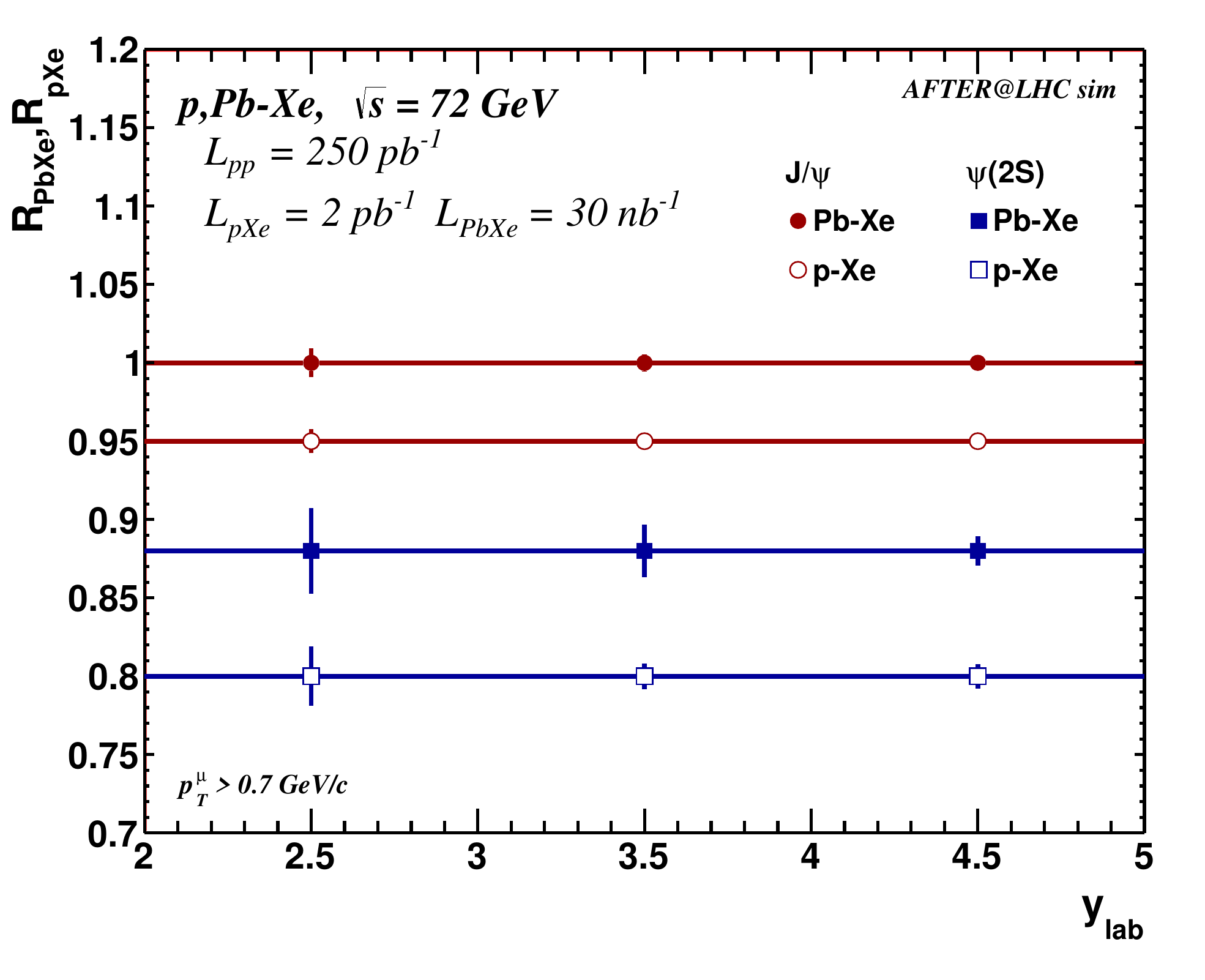}
\end{tabular}
\caption{Left: \jpsi and \psip signals after the (like-sign) combinatorial-background subtraction with the expected statistical uncertainties, \PbXe collisions at $\sqrt{s_{NN}} =$ 72 GeV in 2 $< y_{\rm lab} < $ 5 and $\int \mathcal L_{\rm PbXe}$ = 30 ${ \rm nb}^{-1}$, assuming $R_{AA}=$ 1. Right: Statistical precision of the nuclear modification factors ($R_{{\rm PbXe}}$ and $R_{p{\rm Xe}}$) vs rapidity in \PbXe and \pXe at $\sqrt{s_{NN}} =$ 72 GeV, assuming the combinatorial background subtraction with the like-sign technique. The integrated luminosities are: $\int \mathcal L_{pp}$ = 250 ${\rm pb}^{-1}$, $\int \mathcal L_{p{\rm Xe}}$ = 2 ${\rm pb}^{-1}$, $\int \mathcal L_{{\rm PbXe}}$ = 30 ${\rm nb}^{-1}$, for \pp, \pXe and \PbXe respectively .}
\label{fig:psi_HI}
\end{figure}

The left panel of Figure~\ref{fig:psi_HI} shows the di-muon invariant-mass distribution in the \jpsi and \psip mass region and in the entire rapidity range of 2 $< y_{\rm lab} < $ 5. As in the $\Upsilon$ case, the combinatorial background is subtracted using the like-sign technique. In addition, the contributions from correlated \cc, \bb and Drell-Yan backgrounds are included. The \cc correlated background dominates over the other background sources, under the \jpsi and \psip signal peaks. 

The right panel of Figure~\ref{fig:psi_HI} shows our predicted statistical precision for the nuclear-modification factors ($R_{{\rm PbXe}, p{\rm Xe}}$) for \jpsi (red markers) and \psip (blue markers) in \PbXe (filled symbols) and \pXe (open symbols) collisions at $\sqrt{s_{NN}}$ = 72~GeV. The statistical uncertainties include that of the \pp baseline uncertainties of the \jpsi and \psip expected yields and of the like-sign combinatorial background subtraction - which is the dominant source here. $R_{{\rm PbXe}, p{\rm Xe}}$ can be extracted as a function of rapidity, in three $y$ ranges, with a very good precision.

Such measurements of quarkonia in wide rapidity ranges will allow one to probe the hot and dense medium resulting from these heavy-ion collisions at different temperatures.
Furthermore, to gain even more information on the medium properties, such high \jpsi and \psip expected yields would be used for analyses of $R_{AA}$ differential in centrality, as well as measurements of the elliptic flow, $v_{2}$, which we leave for a future publication.

\subsection{Drell-Yan pair production in heavy-ion collisions at $\sqrt{s_{NN}} =$ 72~GeV }

The physics program of AFTER also includes the precise measurement of the Drell-Yan pair-production process which can probe initial-state effects on quarks using different \PbA collision species. 

\begin{figure}[htp]
\centering
\begin{tabular}{ll}
  \includegraphics[width=0.45\textwidth]{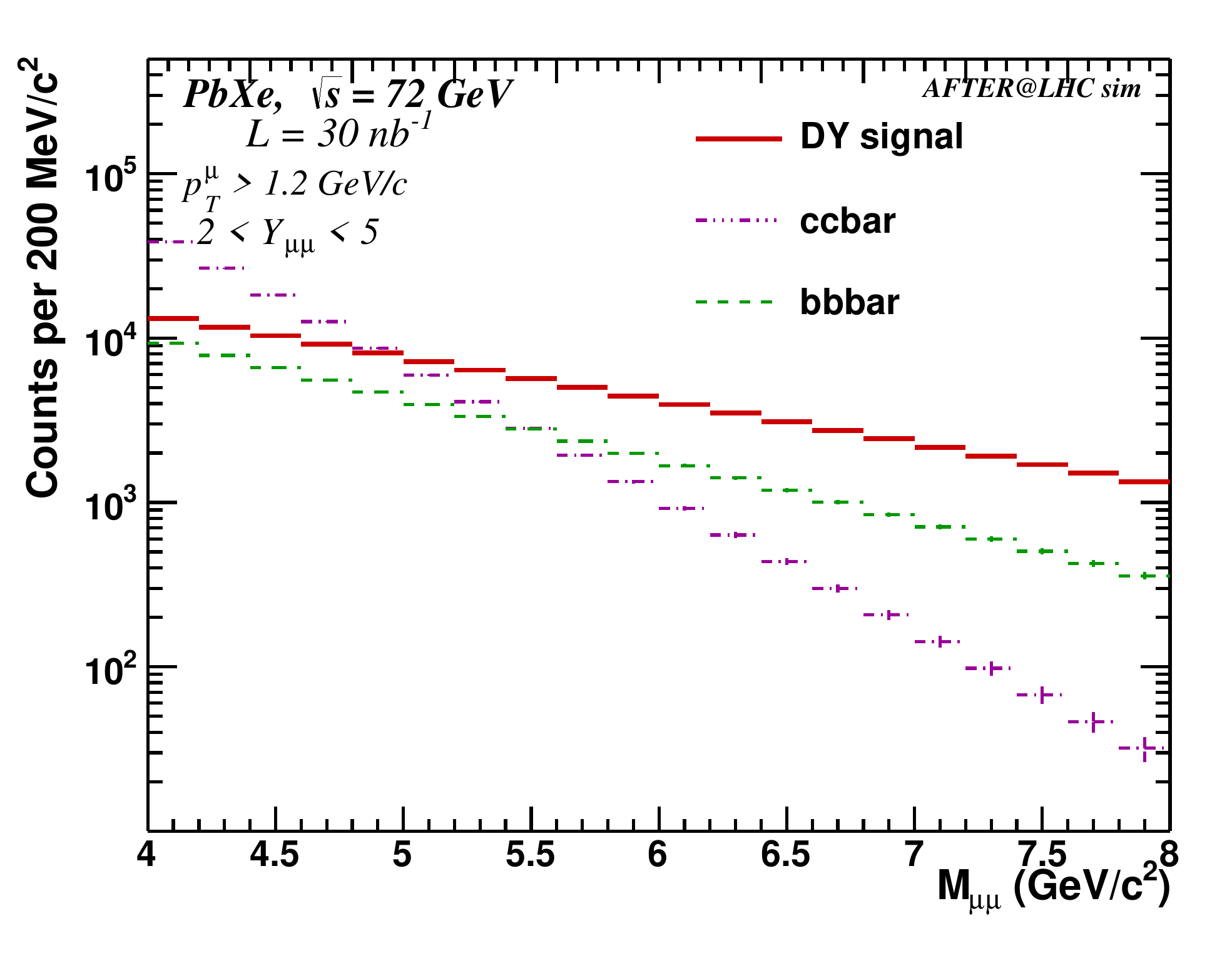}
  \includegraphics[width=0.45\textwidth]{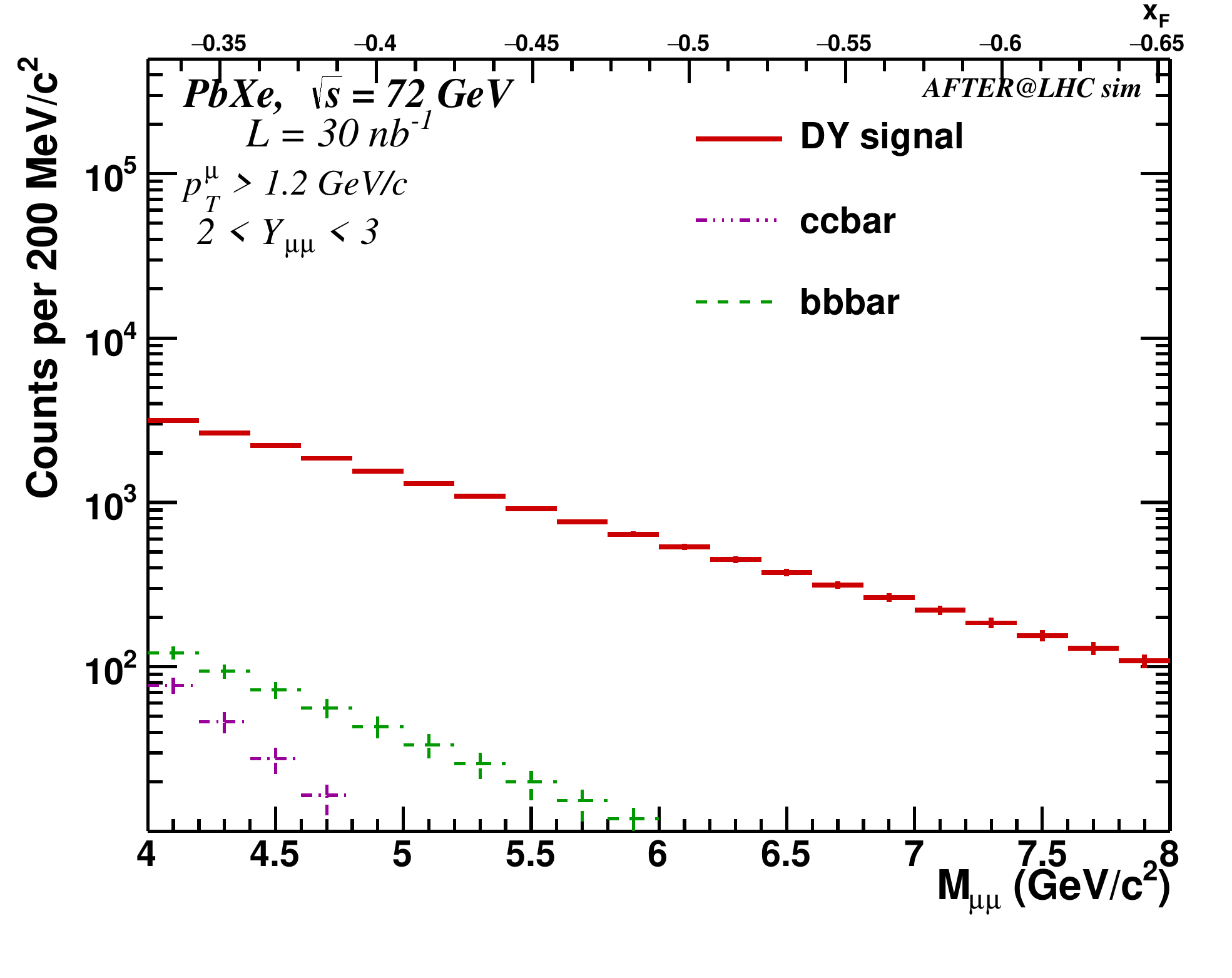} \\
  \includegraphics[width=0.45\textwidth]{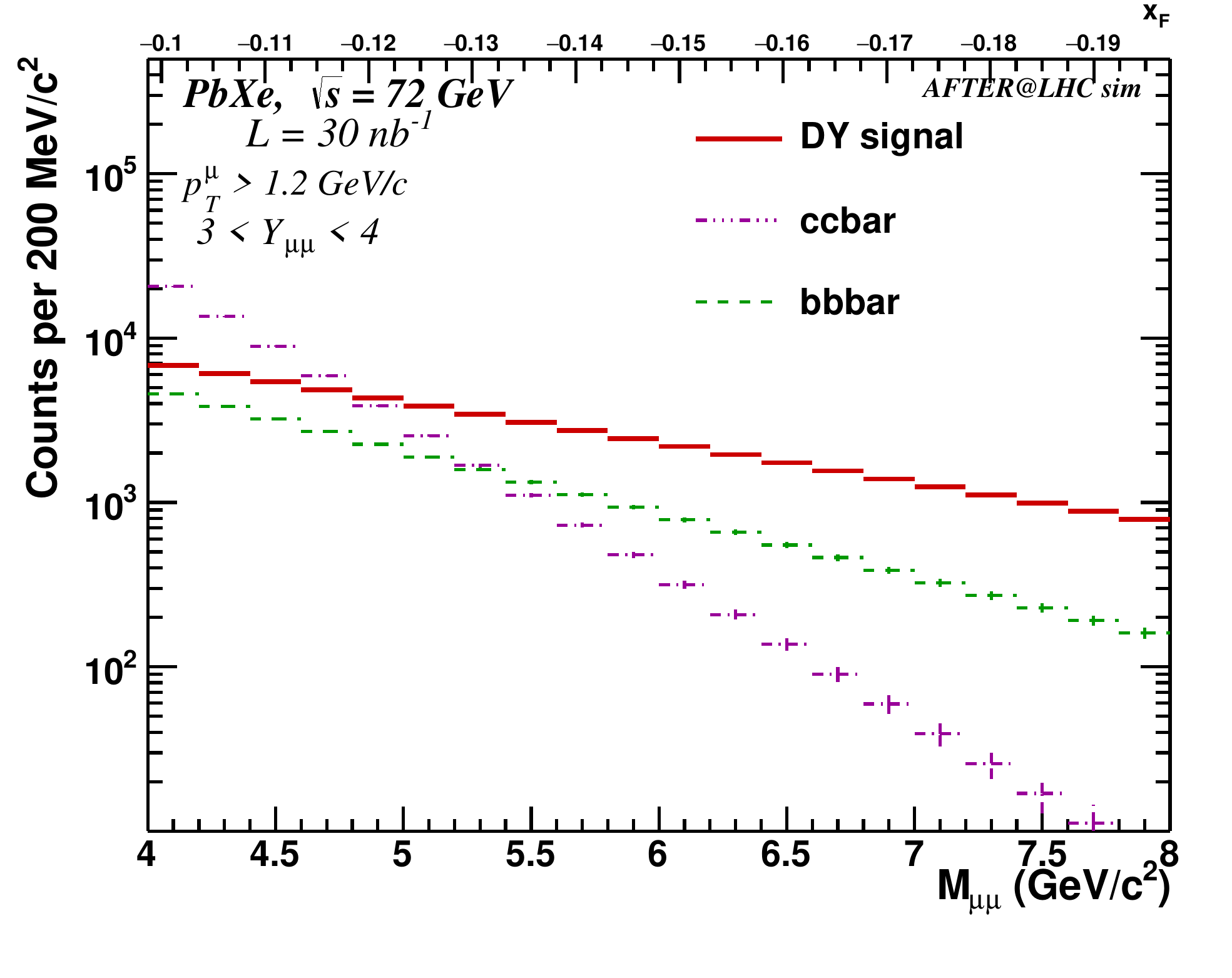}
  \includegraphics[width=0.45\textwidth]{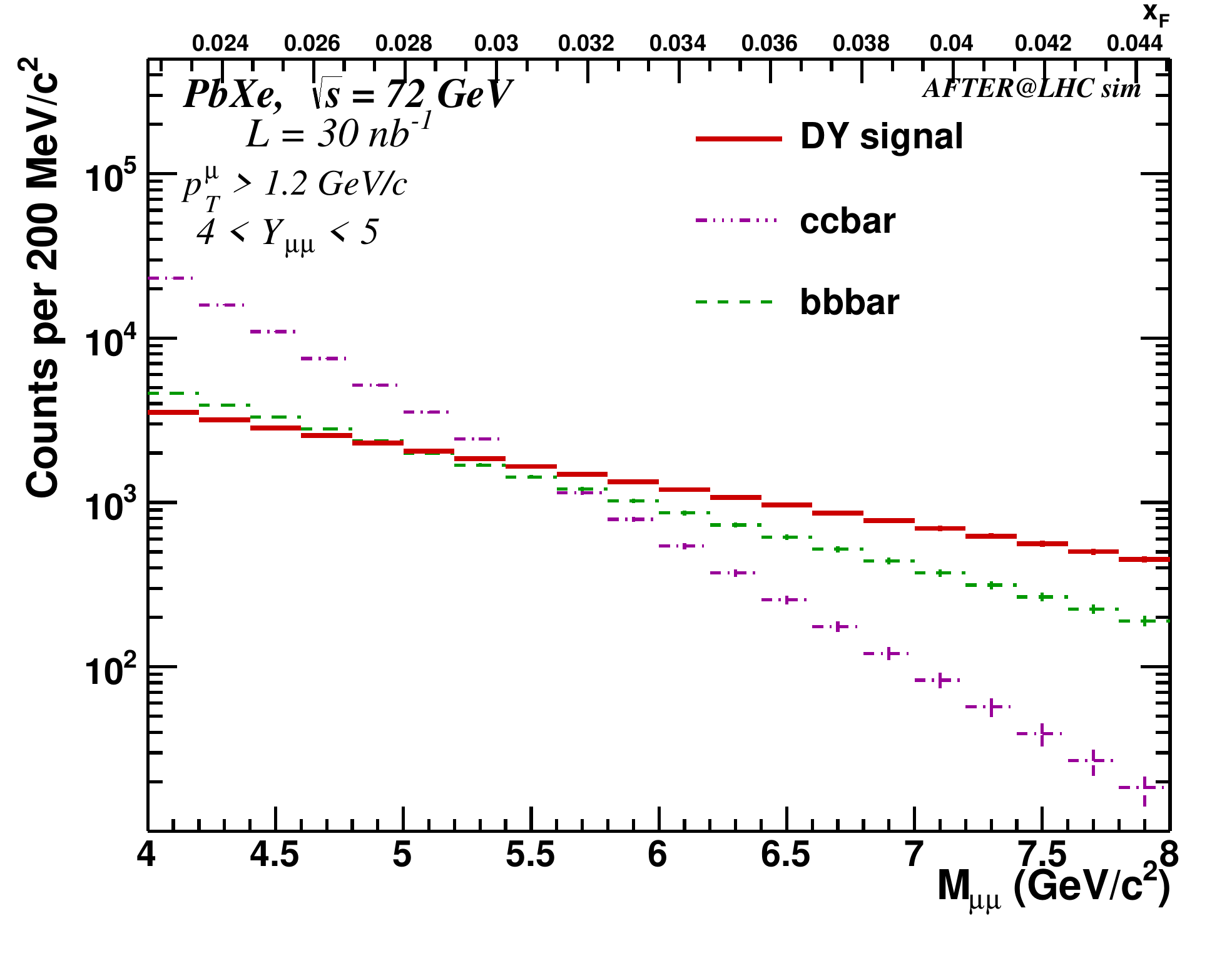}
\end{tabular}
\caption{Di-muon invariant-mass distributions (4 $< M_{\mu^+\mu^-} <$ 8 GeV/$c^2$) for Drell-Yan, \cc and \bb productions, in the integrated rapidity range of 2 $< y_{\rm lab} < $ 5 (top left) and divided into following ranges: 2 $< y_{\rm lab} < $ 3 (top right), 3 $< y_{\rm lab} < $ 4 (bottom left) and 4 $< y_{\rm lab} < $ 5 (bottom right). The upper $x$-axis represents the corresponding $x_{F}$ values in a given rapidity range and invariant-mass bin. The combinatorial background is not presented and systematic uncertainties 
resulting from the background subtraction with with the event-mixing technique are not included. 
\PbXe collisions at $\sqrt{s_{NN}} =$ 72 GeV with $\int \mathcal L_{\rm PbXe}$ = 30 ${\rm nb}^{-1}$, assuming $R_{AA}=$~1.}
\label{fig:DYmass_HI}
\end{figure}

A great challenge in the Drell-Yan measurement at high energies is the large correlated background from $c$+$\bar{c} \rightarrow D^++D^- \rightarrow l^+l^-$ and $b$+$\bar{b} \rightarrow B^++B^- \rightarrow l^+l^-$. This background is smaller at the c.m.s. energy of \AFTER in \PbA collisions, and can easily be further reduced with a modern vertex detector by applying a secondary vertex cuts on displaced production of $D$ and $B$-mesons. Figure~\ref{fig:DYmass_HI} separately shows the expected yields for Drell-Yan, \cc and \bb production in \PbXe collisions as a function of the di-muon invariant mass in the range between the $\psi$ and $\Upsilon$ (4 $< M_{\mu^+\mu^-} <$ 8 GeV/$c^2$), for different rapidity ranges within 2 $< y_{\rm lab} < $ 5. A single muon $p_{T}^{\mu}$ cut of 1.2 GeV/$c$ is applied. We also stress here that going to the very backward rapidity region (2 $< y_{\rm lab} <$ 3) -- which is probably the most interesting one physics wise -- renders the Drell-Yan signal even cleaner since the quark-induced processes are favoured and the background is reduced accordingly.

Due to the continuous nature of the signal di-muon invariant-mass distribution shape and to the reduced $S/B$  ratios, the like-sign technique does not provide enough statistical precision for the signal extraction in this case.
However, one can precisely determine the high combinatorial background expected in \PbA collisions with the event-mixing technique\footnote{where an arbitrarily large number of $\mu^+$ and $\mu^-$ from different events can be mixed and paired}. This technique combined with a large amount of like-sign di-muon background pairs (used to normalise the shape determined by the mixed events) should provide a robust subtraction of the combinatorial background even for small $S/B$ values.
These expected Drell-Yan yields and $S/B$  ratios are gathered in Table~\ref{tab:DY_yields}.
In such case, the main remaining uncertainty would be of systematical origin, thus difficult to simulate. 
As such, the combinatorial background is not accounted for in the distributions shown in Figure~\ref{fig:DYmass_HI}, which should therefore be considered as idealised. On the way, we note that the event-mixing technique can also be applied to improve the extraction of the quarkonium-signal mentioned above.

\begin{table}[!htb]
\caption{Drell-Yan yields and ratios of the yields over the combinatorial background for \PbXe collisions at $\sqrt{s_{NN}} =$ 72 GeV and for $\mathcal L_{\rm PbXe}$ = 30 ${\rm nb}^{-1}$, in 4 di-muon invariant mass ranges between 4 and 8 GeV/$c^{2}$ and 3 rapidity ranges between 2 and 5, with a single muon cut of $p_{T}^{\mu} >$ 1.2 GeV/$c$ on both muons. }
\label{tab:DY_yields}
\begin{center}{\renewcommand{\arraystretch}{1.5}
\begin{tabular}{c|c|c||c|c}
\multirow{2}{1.5cm}{Yields} & \multicolumn{2}{c||}{$M_{\mu^+\mu^-}$: 4-5 GeV/$c^{2}$} & \multicolumn{2}{c}{$M_{\mu^+\mu^-}$: 5-6 GeV/$c^{2}$}
  \\ \cline{2-5}
   & signal ($\times 10^{3}$)  & $S/B$ ($\times 10^{-3}$)
     & signal ($\times 10^{3}$) & $S/B$ ($\times 10^{-3}$)  \\
\hline \hline
$y$: 2-3 & 11.5 	& 2.6	& 4.8	& 6.3   \\ \hline
$y$: 3-4 & 26.9 	& 0.2 	& 15.6 & 0.4  \\ \hline
$y$: 4-5 & 15.3 	& 0.3 	& 8.3	& 1.0 \\
\end{tabular}
\begin{tabular}{c|c|c||c|c}
\multirow{2}{1.5cm}{} & \multicolumn{2}{c||}{$M_{\mu^+\mu^-}$: 6-7 GeV/$c^{2}$} & \multicolumn{2}{c}{$M_{\mu^+\mu^-}$: 7-8 GeV/$c^{2}$}
  \\ \cline{2-5}
   & signal ($\times 10^{3}$)  & $S/B$ ($\times 10^{-3}$)
     & signal ($\times 10^{3}$) & $S/B$ ($\times 10^{-3}$)  \\
\hline \hline
$y$: 2-3 & 1.9 & 13.9 & 0.8 & 28.1   \\ \hline
$y$: 3-4 & 8.8 & 0.9 & 5.1 & 2.1 \\ \hline
$y$: 4-5 & 5.1 & 3.0 & 2.8 & 6.3 \\
\end{tabular}}
\end{center}
\end{table}

\begin{figure}[htp]
\centering
\begin{tabular}{ll}
  \includegraphics[width=0.65\textwidth]{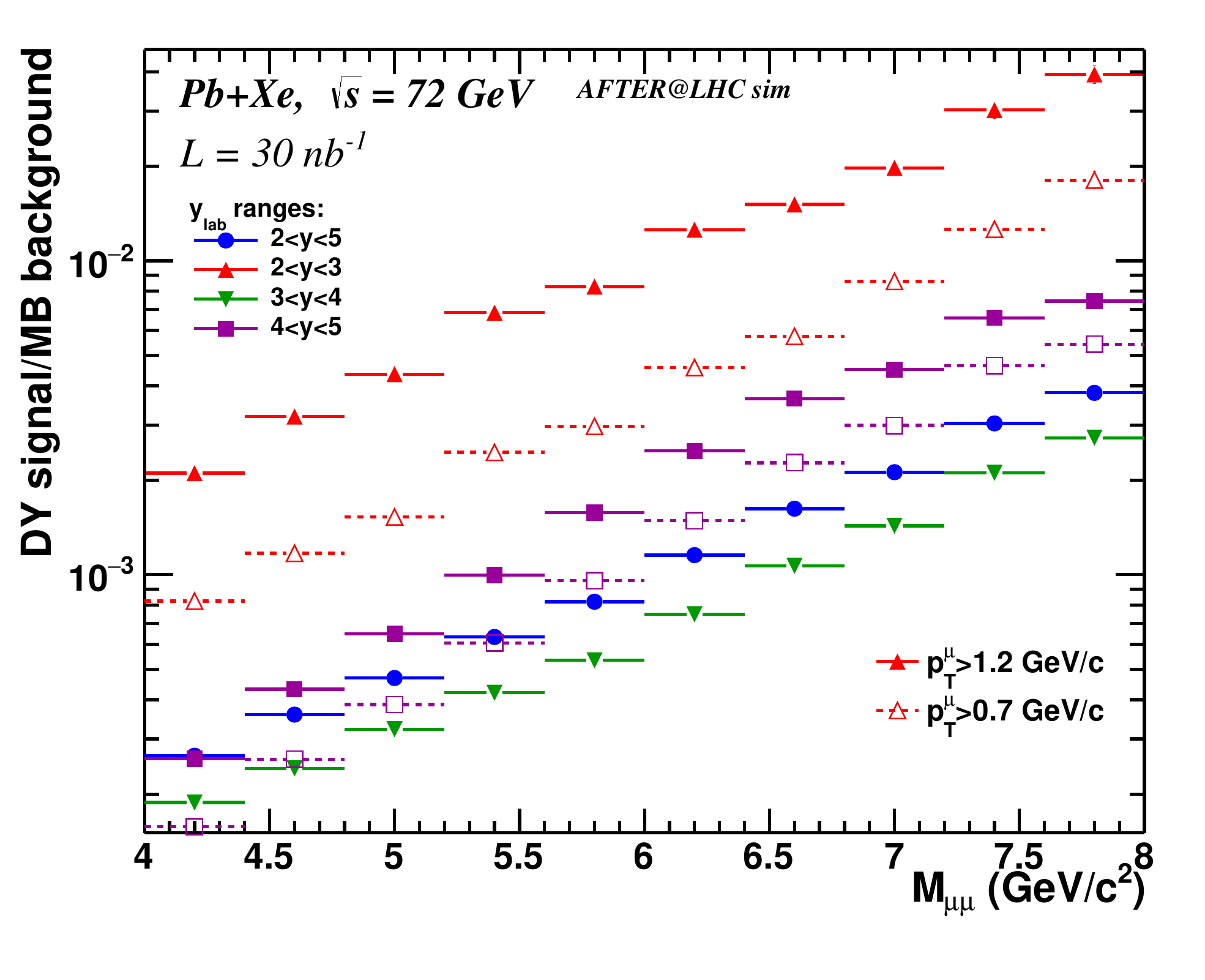}
\end{tabular}
\caption{Drell-Yan signal over the combinatorial background ratio as a function of di-muon invariant mass, for 2 $< y_{\rm lab} < $ 5, 2 $< y_{\rm lab} < $ 3, 3 $< y_{\rm lab} < $ 4 and 4 $< y_{\rm lab} < $ 5. Solid lines represent the $S/B$ with the default Drell-Yan single muon cut of $p_{T}^{\mu} >$ 1.2 GeV/$c$, and dashed lines represent the $S/B$ with $p_{T}^{\mu} >$ 0.7 GeV/$c$.  No nuclear modifications assumed. \PbXe collisions at $\sqrt{s_{NN}} =$ 72 GeV with $\int \mathcal L_{\rm PbXe}$ = 30 ${\rm nb}^{-1}$.}
\label{fig:DYSB}
\end{figure}

We note that the $S/B$ also improves with increasing $M_{\mu^+\mu^-}$ and when going to the most backward rapidity region. This is clear from Table~\ref{tab:DY_yields} and Figure~\ref{fig:DYSB}, where the signal over the combinatorial-background ratio is studied as a function of the di-muon invariant mass for the integrated rapidity region of 2 $< y_{\rm lab} < $ 5, and for the following 3 ranges: 2 $< y_{\rm lab} < $ 3, 3 $< y_{\rm lab} < $ 4 and 4 $< y_{\rm lab} < $ 5. As can be also seen on Figure~\ref{fig:DYSB}, the S/B can be increase by increasing the $p_{T}$ cut on a single muon. In order to maximise the $S/B$, the single muon minimum $p_{T}^{\mu}$ cut is changed from 0.7 GeV/$c$ to 1.2 GeV/$c$. Indeed the particles that contribute to the combinatorial background are mostly produced with lower $p_{T}$. However, a further increase of the minimum muon $p_{T}^{\mu}$ starts to impact the Drell-Yan signal. 

\subsection{Acceptances of the ALICE detectors in the fixed-target mode} 

The ALICE experiment is designed to cope with the high multiplicity of heavy-ion collisions and a fixed-target mode can also be implemented for this experiment. The ALICE detector consists in a central barrel part, which measures hadrons, electrons, and photons and in a forward muon spectrometer~\cite{Aamodt:2008zz,Abelev:2014ffa}. The ALICE upgrade is planned after the Long Shutdown 2 (LS2) of 2018~\cite{Abelevetal:2014cna} and is designed for recording data with a rate of 200 kHz in \pp and \pA collisions and 50 kHz in \PbPb collisions. 

The central barrel covers the pseudorapidity range $|\eta|<0.9$ around mid-rapidity, while the muon spectrometer covers the forward pseudorapidity range $2.5 < \eta < 4$. It should be noticed that, due to the absorber system the only available region for the installation of a fixed target in ALICE is the A-side, opposite to the muon spectrometer one. 

\subsubsection{Muon Spectrometer}

In a fixed-target mode, the acceptance of the muon spectrometer allows for
measurements in the rapidity regions of $-2.3 < \ycms < -0.8$ with a $7$ TeV proton beam and of $-1.7 < \ycms < -0.2 $ with a $2.76$ A.TeV Pb beam, when considering the target at the nominal interaction point of the experiment.  

\begin{figure}[htbp] 
  \begin{center}
  \vspace{-0.1cm}
     \includegraphics[width=0.5\textwidth]{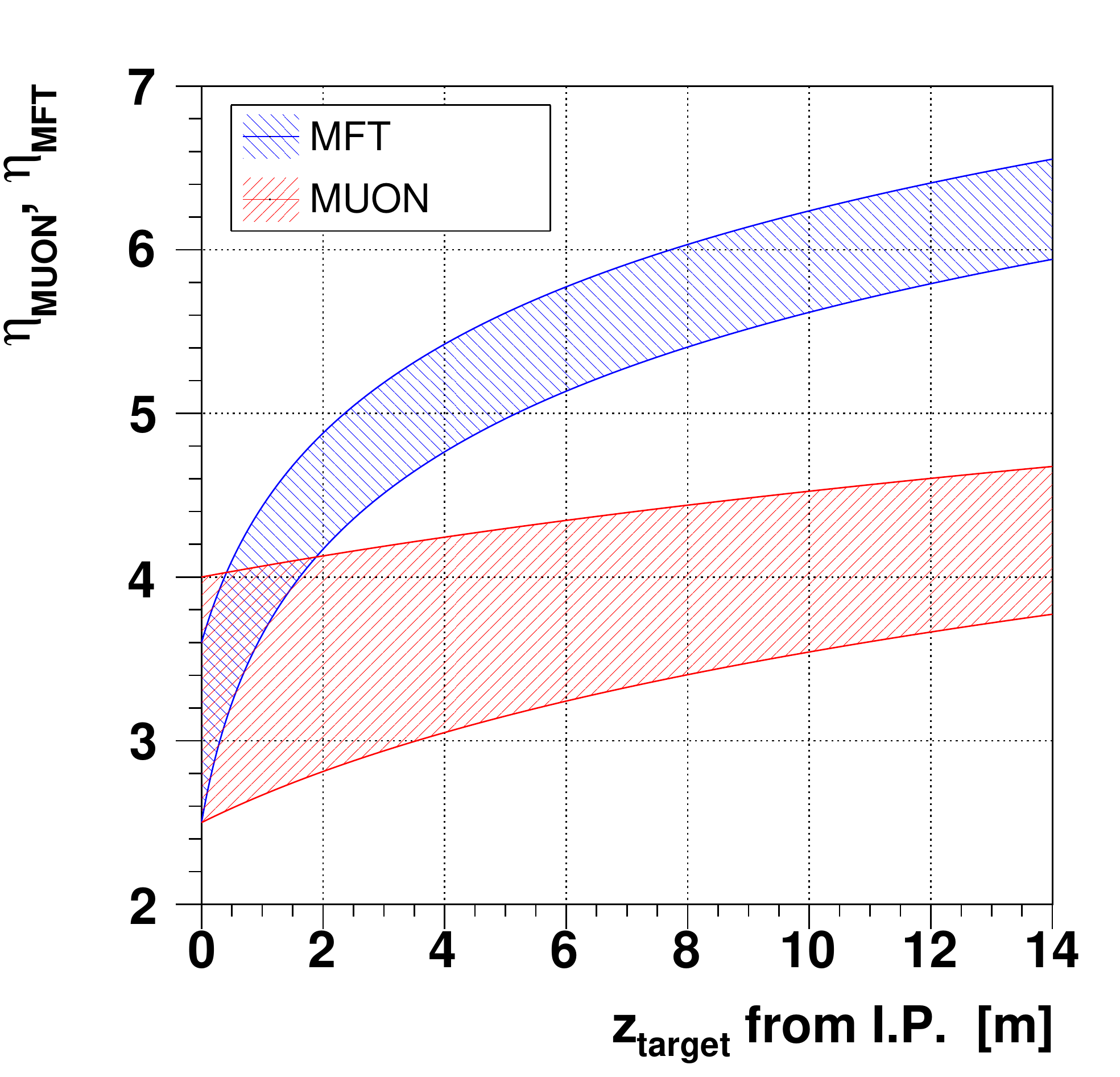}
   \end{center} 
   \vspace{-0.4cm}
   \caption[\textwidth]{Pseudorapidity ranges in the laboratory frame for the tracking chambers of the muon spectrometer (MUON) and the disks of the Muon Forward Tracker (MFT)
   as a function of the longitudinal position of the fixed target from the nominal interaction region. }
   \label{fig:acceptance}
   \vspace{0.2cm}
 \end{figure}

The vertex can in principle be displaced upstream (in the A-side of ALICE) from the nominal interaction point by up to a few meters. The geometrical acceptance of the muon spectrometer as a function of the longitudinal position of the fixed target, is shown in \figurename~\ref{fig:acceptance} separately for the tracking system of the muon spectrometer (identified as
MUON, corresponding to the tracking stations installed after the hadron absorber) and the MFT (Muon Forward Tracker) which will serve as vertex tracker for the muon spectrometer after LS2~\cite{CERN-LHCC-2015-001}. 

As one can notice from \figurename~\ref{fig:acceptance}, the geometrical acceptances of the spectrometer and the MFT decouple very early as the target is moved away from the nominal interaction point. For this reason, an optimised inner-tracking system would be needed, providing both vertexing capabilities and the proxy for the tracking in the spectrometer.

On the other hand, the geometrical acceptance of the spectrometer stays reasonably close to $\ycms \simeq 0$, keeping a range of about one unit of rapidity even for longitudinal positions of the fixed target very far from the nominal interaction point.

It should be however noted that large displacements between the fixed-target position and the hadron absorber would automatically result in an increase of the combinatorial background from semi-muonic decays of pions and kaons. This aspect should be carefully accounted for when evaluating the physics performance of the muon spectrometer in a fixed-target configuration with displaced vertex.

\subsubsection{Central Barrel}

If the target is located at the nominal interaction point of the experiment, the central barrel covers the far backward region with $\ycms<-3.9$ and $\ycms<-3.3$ with a $7$ TeV proton beam and a $2.76$ A.TeV Pb beam, respectively. The detection and the identification of particles with the central barrel therefore seems very appealing to access far negative $x_F$, down to the target fragmentation region. Several questions however need to be addressed in the case of a target far from the nominal interaction point. The tracking performance of the Time Projection Chamber, Time Of Flight and Transition Radiation Detector need to be evaluated for particles entering and traversing their volume with an angular direction different from the nominal specifications. 

As in the muon spectrometer case, a new vertex detector would be needed if the current ALICE Inner Tracking System (ITS) falls outside the geometrical acceptance of interest for particle production in a fixed-target configuration. It would provide both vertexing capabilities and the proxy for the tracking in the main detectors of the central barrel.

\section{Conclusions}
We have outlined the broad spectrum of possibilities offered by \AFTER for studies of the nuclear matter properties in \pA and heavy-ion collisions in an energy range where a phase transition to a partonic matter in heavy-ion collisions is expected. 

The heavy-ion studies with the LHC beams used in the fixed-target mode include:
\begin{itemize}
 \item the correlation of the medium temperature and the quarkonium suppression as a function of their binding energies by looking for signs of a sequential suppression with as many states as possible;
\item the precise and complete determination of the Fourier expansion of identified particle azimuthal asymmetries as a function of rapidity, down to the target fragmentation region;
  \item the determination of different mixtures of quark-gluon plasma and cold nuclear matter effects on the rapidity- and transverse-momentum-dependent particle-yield modifications, in particular heavy-flavoured particles;
  \item the verification/falsification of the factorisation of the initial-state nuclear effects on Drell-Yan-pair measurements on different systems including asymmetric nucleus-nucleus collisions;
    \item the search for net baryon singularities from a possible QCD critical point using the beam-energy-ramp phase;
  \item the search for modifications in the mass width and branching ratios between hadronic and leptonic channels of $\phi$, $\omega$ and $\rho$ decays as a signature of the chiral-symmetry restoration in the same spirit of the NA60 studies.
\end{itemize}

As we have shown for a LHCb-like detector, various charmonium and bottomonium states can be measured with an unprecedented accuracy over wide \pT\ and rapidity ranges both in \pA and \AA collisions. These studies will be accompanied by open heavy-flavour, light-flavour, and Drell-Yan measurements to deliver in-depth characteristics of the partonic matter and the phase transition. Based on our first acceptance studies, the ALICE detectors used in the fixed-target mode can provide complementary measurements in this new energy domain.

\section*{Acknowledgements}
This research was supported in part by the French P2IO Excellence Laboratory and the French CNRS via the grant FCPPL-Quarkonium4AFTER, and Ministerio de Ciencia e Innovacion of Spain and Xunta de Galicia. 

\bibliographystyle{utphys}

\bibliography{AFTER-NewObservables-HIC}

\end{document}